\documentclass[aps,preprint,floatfix,superscriptaddress]{revtex4-1}
\usepackage{graphicx}
\usepackage{amsmath,amsthm}
\usepackage{amssymb}
\usepackage{braket}
\usepackage{xcolor}
\usepackage{bm}
\usepackage{ulem}
\usepackage[bookmarks=false,linkcolor=blue,urlcolor=blue,colorlinks,citecolor=blue]{hyperref}
\newcommand{\be}[1]{ \begin{eqnarray} \mbox{$\label{#1}$} }

\newcommand{\ee}{\end{eqnarray}}

\newcommand{\eeq}{\end{equation}}

\newcommand\ie {{\it i.e. }}

\newcommand\half{\frac 1 2 }

\newcommand{\bracket}[2]   {  \left<#1 |  #2\right>}

\newcommand\oncite [1] {Ref.\,\onlinecite{#1} }
\newcommand\oncitep [1] {Ref.\,\onlinecite{#1}. }
\newcommand\oncitec [1] {Ref.\,\onlinecite{#1}, }

\newcommand{\sech}{\text{sech}}
\newcommand{\csch}{\text{cosech}}

\begin{document}
\pacs{xxx,xxx,xxx  }
\title{Dynamics of ``Classical" Bosons, Fermions, and beyond}

\author{Varsha Subramanyan}
\affiliation{Theoretical Division, Los Alamos National Laboratory, Los Alamos, NM USA}
\author{T.H. Hansson}
\affiliation{Department of Physics, Stockholm University, AlbaNova University Center, 106 91 Stockholm, Sweden}
\author{Smitha Vishveshwara}
\affiliation{ Department of Physics, University of Illinois at Urbana-Champaign, Urbana, IL USA}

\begin{abstract}

We study the classical mechanics and dynamics of particles that retains some memory of quantum statistics. Our work builds on earlier work on the statistical mechanics and thermodynamics of such particles. Starting from the effective classical manifold associated with two-particle bosonic and fermionic coherent states, we show how their exchange statistics is reflected in the symplectic form of the manifold. We demonstrate the classical analogues of exclusion or bunching behavior expected in such states by studying their trajectories in various quadratic potentials. Our examples are two-particle coherent states in one dimension and two-particle vortex motion in the lowest Landau level. We finally compare and contrast our results with previous simulations of the full quantum system, and with
existing results on the geometric interpretations of quantum mechanics.
\end{abstract}
                                        
\maketitle                 

\section{Introduction}  
In this paper we analyze the dynamics of what might sound as an oxymoron, namely ``classical fermions" and  ``classical bosons", and more generally, ``classical anyons". These are particles moving in one or two space dimensions that are described by a classical Lagrangian with a conventional Hamiltonian, but with modified Poisson brackets that emulate aspects of quantum statistics. The statistical mechanics and thermodynamics of such particles was analyzed in \oncitec{hill};  here it was shown that the statistics parameter inherited from the quantum dynamics describes the reduction of the phase space volume, which in turn implies changes in the statistical mechanics and thus the thermodynamics. These approaches exhibit clear similarities with Haldane's concept of exclusion statistics \cite{Haldaneexcl}, though at present we lack techniques  for  extending the description in \oncite{hill} to higher dimensions. 

As stated in \oncite{hill} the  ``statistical interaction" does not amount to any force acting on the particles, but the conclusion that it does not affect the trajectories is in general not true. For instance, explicit examples of how the statistics affect trajectories defined by coherent states are given in an earlier paper of two of us \cite{b4}. In this work we explore how such statistical effects are manifested in the classical dynamics of the theory developed in \oncite{hill}.  As a key observation, we show by a simple example that classical particles, subject to the same forces, \ie having identical hamiltonians, follow different trajectories and move with different speed depending on the ``classical statistics". Our treatment therefore offers a generalized formulation for quantum many-particle dynamics in which classical equations of motion encapsulate some effects of quantum statistics.

This formulation enables studying dynamics in a range of situations and pinpoints the marked effect of quantum statistics. As in \oncite{hill}, our studies focus on two physical systems--particles in 1d  and vortex motion in 2d, in particlar charged particles in the lowest Landau level (LLL). We exemplify both cases with quadratic hamiltonians, and  both cases show fundamental differences between fermions and bosons (which we refer to henceforth as classical fermions and bosons) through the appropriate symplectic forms that affect the corresponding classical equations of motion. The 1d situation offers a palpable instance for gleaning statistical effects. Not only does it demonstrate the intuitive repulsive or attractive behavior that one associates with quantum statistics, but also encapsulates the non-intuitive quantum content carried by commutation rules.  The 2d system shows a richness of behavior in both unbound and bound situations. Saddle potential geometries offer a 2d counterpart to  the 1d inverted harmonic potential in that they are unbounded quadratic potentials with equipotential trajectories such that in both cases small changes in the initial conditions give vastly different asymptotic trajectories. Elliptical potential traps reveal the unique ways in which fermions and bosons avoid or attract each others, and instead of following closed orbits, densely cover energetically allowed regions.

Since the late 70's, it has been known that the arguments based on many-particle wave function properties for having only two types of exchange statistics do not hold in one or two spatial dimensions\cite{Leinaas1977}. 
In one dimension, exchanges always imply collisions. So the truly interesting case comes down to two dimensions where the quantum statistics can be separated from interactions. This is most easily seen in the path integral approach, where one can view exchanges of identical particles as world lines forming a braid, and assign a ``statistical phase factor" to any set of trajectories, by giving a factor $e^{-i\theta}$ to all elementary clockwise exchanges of two adjacent particles. For bosons and fermions, the angular parameter $\theta$ is 0 or $\pi$,  respectively. But $\theta/\pi$ can take any  value, and the corresponding particles as called anyons\cite{PhysRevLett.49.957}. Since non-trivial braids do not exist in dimensions higher than two, anyons can only appear in one or two dimension. The study of anyons as fractionalized excitations over topologically non-trivial many-body ground states has grown to include several sophisticated theoretical and computational approaches, with their recent experimental detection in quantum Hall systems only adding to the excitement\cite{STERN2008204, Greiter, Nakamura20,Bartolomei20}.


Our studies of vortex motion in 2d have direct application to the recent interest in anyons in the quantum Hall systems as well as broader issues concerning its fundamental quasiparticle excitations. The vortex degrees of freedom  can describe charged quantum particles in the lowest Landau level\cite{Kjonsberg,Hansson}. While most experiments manipulate edge-state anyons, as is becoming more evident, their full 2d physics in bulk landscapes is important, with the saddle potential being highly relevant to beam-splitter setting and trapping potentials to interferometer settings\cite{Saddle2,b3}. Most of our examples will contrast the behavior of pairs of fermions and bosons with that of distinguishable particles. The generalization to systems of two anyons is straightforward. In our treatment, we provide the initial steps involving symplectic factors and the associated machinery for analyzing "classical anyons".  The full analysis, however, is computationally more demanding and does not give any qualitatively new insights compared to fermions and bosons. We shall also study three-particle systems, and in this case, there are no exact solutions for anyons.


In Section \ref{background} we provide sufficient background material, mainly from \oncite{hill}, to make this article reasonably self contained, and also introduce the necessary notation. 
In Section \ref{1dho} we use the example of an inverted harmonic oscillator to illustrate the different behavior of bosons and fermions and describe these differences qualitatively in terms of phase space repulsion. 
We devote Section \ref{lllcs} to the study of trajectories of particles in the LLL subjected to various quadratic potentials. In Section \ref{chaos} we provide some evidence, based on numerical estimate of Lyaponov exponents, for possible deterministic chaos in a system of three classical fermions or bosons moving in an elliptical potential, something which is not possible for distinguishable particles. Finally,  in Section \ref{quantcomp} we compare and contrast these results with insights from the geometric interpretation of quantum mechanics. We end with a section on discussions and conclusions.


\section {Background}     \label{background}

Quantum mechanical particles differ from classical ones in two fundamental ways. Firstly, they can behave both as particles and waves, the connection being the Born rule that states that the amplitude squared of the the wave function for a particle, $|\psi(\vec r)|^2$, is the probability amplitude for finding the particle at position $\vec r$. 

Secondly, particles which share {\it all} characteristics, such as mass, spin and charge, are {\it identical}, which means that the probability amplitude of a many particle wave function must be symmetric under interchange which implies that $|\psi(\vec r_1, \vec r_2 \dots \vec r_i\dots \vec r_j \dots \vec r_N|^2 $\\$= |\psi(\vec r_1, \vec r_2 \dots \vec r_j\dots \vec r_i \dots \vec r_N|^2$. It was understood early on that this property implies that the wave function itself has to be either fully symmetric, or fully anti-symmetric under exchange of identical particles. It was later recognized that Lorentz invariance implies that these two classes differ in that the particles with symmetric wave functions, called bosons, all have integer spin, while the fermions with anti-symmetric wave functions all have half integer spin, and this is fully consistent with all observations.

The notion of identical particles has fundamental consequences for statistical mechanics and thermodynamics, and was, in a sense, already  implicit in the classical theory, where the Gibbs paradox could be avoided by assuming particles to be identical so that micro states differing only by a permutation should not be counted separately in the partition function. Also, to calculate entropies from statistical mechanics correctly,  one had to introduce a minimal cell size, $h$, in phase space;  in quantum theory, this is precisely Planck's constant $h$, which also enters the dynamical equations via the canonical commutation relations $[x,p] = i\hbar$.  

It is, at least heuristically, understood how classical dynamics, \ie deterministic particle trajectories, emerge from a quantum mechanical description, either by finding extremal paths in a path integral, or by forming quasi-classical wave functions (coherent states) and calculating the time evolution of the expectations value of the position operator in the limit of small $\hbar$. This way of taking the classical limit however does not make any distinction between bosons and fermions. In particular, there is no classical counterpart to the Fermi exclusion principle, which is essential for understanding the stability of matter.
In this paper we shall consider the alternative way, described in \oncite{hill}, of taking the classical limit  where a memory of the quantum statistics is left in the classical physics.


There are  aspect of quantum statistics that is  directly related phase space properties, and thus statistical mechanics. The Fermi exclusion principle, that prevents more than one fermion to occupy the same cell in phase space, is well known. Haldane later extended this `` exclusion statistics"  to the case of anyons\cite{Haldaneexcl}. For bosons there is, to our knowledge, no such clear-cut consequence of statistics, but rather vague statements like ``bosons like to be in the same quantum state", often in  connection to BE condensation, or photon bunching in photonics. However,  detailed studies of coherent states of pairs of bosons and fermions, show sign a ``statistical attraction" between bosons\cite{b1,b2,b3,b4,b5}. 

A while back, it was realized that the exclusion statistics has a classical counterpart if the classical limit is taken in an appropriate way, as described in \oncitep{hill} The approach in that work was to start from a coherent many-particle wave function that appropriately takes into account the quantum statistics, and depends parametrically  on the particle positions $\vec r_i$, which can be thought of as the centers of coherent state wave-packets. From this, one can extract a classical phase space actions for the $\vec r_i$, where the Hamiltonian is the expected one, while the symplectic form, which defines the classical brackets is modified and distinguishes bosons from fermions. \oncite{hill} also showed that the quantum statistics parameter $\nu$ is related to the excluded classical phase space volume $\alpha$, corresponding to adding an extra particle, by $\alpha = \nu h$. Thus, to maintain an exclusion effect in the classical theory, one must take a double limit where $\alpha$ is kept constant even at small $\hbar$.  That is, the projection mechanism employed here is not a classical limit in the sense of taking $\hbar\rightarrow 0$, but rather an alternative way to define a classical mechanics which retains elements of quantum statistics.  It is also notable that the formulation of the classical limit of quantum statistics can be extended to include anyons and fractional statistics.While this work is concentrated on fermions and bosons, all the results for two particle systems are readily generalized to anyons.



\subsection{Notation and elements of classical mechanics } \label{classmec}

Consider a phase space Lagrangian, 
 \be{fundlag}
 L = \sum_i a_i(q_1 \dots q_{2N})\dot q_i - H(q_1 \dots q_{2N})\, ,
 \ee
 where the $q_i$ are $2N$ coordinates, and $a_i$ the  symplectic potential, or Liouville one-form.  This generalizes the more conventional case of a Lagrangian which is quadratic in time derivatives, and where the first term in Equation \eqref{fundlag} can be  written as $\sum_i p_i\dot q_i$, where $q_i$ and $p_i$ are independent phase space variables. 
To define a Hamiltonian dynamics from  Equation \eqref{fundlag}, we follow the
general method described in Ref. \cite{PhysRevLett.60.1692} for obtaining  the canonical brackets, so that the Hamiltons equations reproduce the correct dynamics. To do this, first define the symplectic field strength by ${\cal F}_{ij} = \partial_{q_i} a_j -\partial_{q_j} a_i  $ and define the Poisson brackets as,
 \be{poisson}
 \{q_i,q_j\}= {\cal F}^{-1}_{ij} \, . 
 \ee
One can now, by direct calculation, verify that the Hamilton's equations obtained from $H$ using this bracket coincide with the Euler-Lagrange equations obtained directly from Equation \eqref{fundlag}. 

 We emphasize that the  $q_i$ in Equation \eqref{fundlag} are  {\it phase space coordinates}, and it is not necessary, or sometimes even possible, to define a configuration space of coordinates and velocities. However, in the simple case of $N$ non-relativistic particles moving on a line, we can take $x_i = q_i$ and $p_i = q_{N+i}$ for $i=1,2,\dots N$,
 therefore, by taking $a_i =  p_i$, \eqref{fundlag} becomes the usual phase space Lagrangian, $L= \sum_i p_i \dot x_i - H(x_i,p_i)$. Using Equation \eqref{poisson}  we then regain $\{ x_i, p_j\}=\delta _{ij}$.
 
 An example more close to what we shall consider in the following is vortex motion which we can conveniently think of as guiding center motion in the lowest Landau level. Here we take 
  $X_i =  q_i$, and  $Y_i =  q_{N+i}$ for $i=1,2,\dots N$ and  $a_i = Y_i /\ell^2$ where $(X_i, Y_i)$ is the coordinates of the vortex center, and $\ell$ some pertinent length scale, so  the bracket becomes $\{ X_i, Y_j\}=\ell^2 \delta _{ij}$. In the case of charged particles in the lowest Landau level (LLL), $X$ and $Y$ are guiding center variables, and $\ell_B$ is the magnetic length.
  
  In the following we shall use the Hamiltonians
 \be{hamilton}
 H_p= \frac 1 {2m} p^2 \pm \half m\omega^2 x^2 \quad; \quad  H_v = \left(\frac X a\right)^2 \pm \left(\frac Y b\right)^2 
 \ee 
  for the harmonic oscillator Hamiltonian $H_p$ in the particle picture, and an elliptic or hyperbolic potential $H_v$ in the vortex picture respectively. Neither of these are bounded from below if we chose the minus sign. 
  
 Below we will mainly use dimensionless complex coordinates defined by
  \be{complex1d}
z &=& \frac x {\ell_{ho}} +  i \frac {\ell_{ho}} \hbar \, p  \quad, \quad \text{particle motion in 1d}  \\
z &=& \frac 1 {\ell_B} (X + iY) \quad , \quad
\text{charged particle in the LLL} \label{complexlll}
\ee
where the two length scales are $\ell_B^2 = eB/\hbar$ and $\ell_{ho}^2 =  \hbar/(m\omega)$ respectively. As was pointed out in \oncite{hill}, the reason that $\hbar$ appear in these, and several of the following,``classical" expressions is that we are using dimensionless coordinates. Any constant with dimension of action could be used. 
 

 \subsection{From a manifold of quantum states to a classical Lagrangian } \label{manifold}

We shall now show recall how to extract a classical theory from a manifold  of quantum states characterized by a set of of set of parameters. 
  
In analogy with Equation \eqref{fundlag}, a general quantum system can be described by way of the Lagrangian,
\be{lag}
    L =i\hbar\braket{\psi|\dot{\psi}}-\bra{\psi}\hat{H}\ket{\psi},
    \ee
where variation with respect to the bra $\bra\psi$ gives the usual Schr\"odinger equation $H\ket\psi = i\partial_t\ket\psi$.

 In many applications, one is not interested in the full Hilbert space, but typically just in some manifold of low energy states. Examples are the lowest Landau level (LLL) states in a strong magnetic field, or the Slater determinants formed by states in the non-empty bands in a metal or an insulator. To study these kinds of situations, one first parametrize the states in the submanifold,  ${\cal M}_q = \{ \ket{\psi_{x}} \}$ of normalized quantum states indexed by the parameters $x = \{x_i\}$ which forms a classical manifold ${\cal M}$. 

 Assuming that the low energy dynamics can be well described by motion restricted to ${\cal M}_q$, the time evolution is coded in $x_i(t)\in {\cal M}$ and Equation \eqref{lag} reduces to 
\be{sympform}    
    L =\dot{x}_iA_i-V({x})\label{Lag} \, ,
\ee
where repeated indices are summed over, and $V(x)=\bra{\psi_x}\hat{H}\ket{\psi_x}$.\footnote{
We use the notation $q_i$ to denote point in the classical manifold $\cal M$, both to adhere to the notation in \oncite{hill}, and to emphasize that these coordinates index quantum states.
}
The symplectic potential $A_i$ is the Berry connection 
\be{berconn}
A_i (x) =i\hbar\braket{\psi_x|\partial_i\psi_x} \, .
\ee
 where we used the simplified notation $\ket{\partial_i \psi_x} \equiv \partial_{x_i} \ket {\psi_x} $ for derivatives with respect to the parameters.   With this we have extracted a classical Lagrangian from the quantum theory, and the time evolution is in the classical manifold, $\cal M$ spanned by the coordinates $x_i$.
 Although there is a one-to-one mapping between ${\cal M}_q$ and $\cal M$, it is useful to keep the distinction, since the elements of the former are quantum states, while in the latter they are classical coordinates.

The Euler-Lagrange equations from Equation \eqref{sympform} read
\be{eom}
    \dot{x}_i =(f^{-1})_{ij}\partial_j V(x)
\ee

where the Berry field strengh tensor $f_{ij}$ is
\be{berfield}
    f_{ij} =\partial_iA_j-\partial_jA_i \, ,
\ee
and the brackets are again given by Equation \eqref{poisson}.
 
Clearly, $A_i$  has a geometric meaning, and it turns out to be fruitful to examine the geometry of the manifold ${\cal M}_q$.  To define a meaningful derivative on ${\cal M}_q$, we must remember that the quantum states are rays in the Hilbert space, so a component along the state does not correspond to a physical change. This leads us to define the projected derivative, 
 \be{projder}
\ket{ D_i  \psi_x} \equiv  (\partial_i + \frac i \hbar A_i) \ket{\psi_x} = \ket{\partial_i \psi_x}  - \ket{\psi_x}\bracket{\psi_x}{\partial_i\psi_x} 
 \ee 
 so that the scalar product of  the covariant vectors $\ket{D_i\psi_x}$ on the tangent space of ${\cal M}_q$ becomes,
 \be{qgt}
 \braket{D_i\psi_x|D_j\psi_x} = \braket{\partial_i\psi_x|\partial_j\psi_x}- \bracket{\partial_i\psi_x}{\psi_x}\bracket{\psi_x}{\partial_j\psi_x}
 \ee
 which is the (quantum) geometrical tensor introduced by Provost and Vallee\cite{proval}.
 It is now straightforward to show that the Berry field strength, can be written as, 
\be{symp}
f_{ij}=-2\hbar\, \mathfrak{Im}\braket{D_i\psi_x|D_j\psi_x} \, ,
\ee
which also defines a symplectic two form, 
\be{stwoform}
    \omega =-\frac{1}{2}f_{ij}dx^i\wedge dx^j
\ee
that determines the phase space volume.

Although Equation \eqref{sympform} looks completely classical, we shall see below that aspects related to the original quantum thoory are retained through a non-trivial dependence of $f_{ij}$ on the quantum  statistics. 

As an aside, we mention that the real part of the scalar product $\braket{D_i\psi_x|D_j\psi_x}$ also has  a geometric interpretation as a metric on $\cal M$, {\it viz.}
\be{metric}
g_{ij} = 2\hbar\, \mathfrak{Re}\braket{D_i\psi_x|D_j\psi_x} \, . 
\ee
This Fubini-Study metric defines distances on $\cal M$.  This aspect of the geometric tensor has been emphazised in the recent literature\cite{torma}, but we shall not pursue this further. 

Since we deal with complex manifolds,  the Hermitian inner product Equation \eqref{qgt} defines $\cal M$ as a Hermitian manifold, and we shall in fact consider a more restricted structure where the symplectic form $\omega$ is closed, \ie $d\omega = 0$. This defines a {\it  K\"ahler manifold}, and $\omega$ can be expressed as $\omega = \frac i 2  \partial_{z_i} \partial_{\bar z_j}\mathbb{K}(\bar z, z) \, d z_i\wedge d \bar z_j $, where $\mathbb{K}$ is the {\it  K\"ahler potential}. (We use the notation $z=x + iy$)\cite{geom1}.  

As detailed below, for the specific examples we shall consider, the states can be written as 
\be{holstat}
\ket {\psi_x} =  N(\bar z, z)\ket{\phi_i(z)}
\ee
where $N$ is a normalization constant and the kets 
$\ket{\phi_i(z)}$ are holomorphic.  A direct calculation gives,
\be{comppot}
 A_{z_i} =- i\hbar\partial_{z_i}\bar N(z_i,\bar{z}_i) \quad , \quad 
 A_{\bar{z}_i} = i\hbar\partial_{\bar{z}_i}N(z_i,\bar{z}_i) \, ,
\ee
and picking the phase convension $N=\bar N$ we get,\,  
\be{holpot}
 A_{z_i} =\frac{i}{2}\partial_{z_i}\mathbb{K}(z_i,\bar{z}_i) \quad , \quad 
 A_{\bar{z}_i} =-\frac{i}{2}\partial_{\bar{z}_i}\mathbb{K}(z_i,\bar{z}_i).
\ee
with 
\begin{align}
    \mathbb{K}(z_i,\bar{z}_i)&=\hbar\ln{|N(z_i,\bar{z}_i)|^{-2}}\, .
\end{align}
Thus $ f_{\bar{z}_iz_j} =\partial_{\bar{z}_i}\partial_{z_j}\mathbb{K}(z_i,\bar{z}_i)$, and the states in Equation \eqref{holstat} indeed span a K\"ahler manifold.

\subsection{Coherent States}\label{CS}
 
We now  use the method described above to derive the classical Lagangian and equations of motion for the   physical systems described in the beginning of this section. We focus on two-particle systems and compare the dynamics obtained from the classical equations of motion for fermionic, bosonic, and distinguishable particles. 

The particle coordinates  are extracted from  coherent quantum states describing the $2N$ dimensional phase space of $N$ quantum particle moving on a line, or being restricted to the LLL. In both cases, the states are ``coherent" in that they have minimum uncertainty in the pertinent phase space. 

Recall that for a single particle the coherent states $\ket z$  are eigenstates of an annihilation operator $a$, that is, 
\begin{align}
a\ket{z}=z\ket{z} \label{eq:cs} \, ,
\end{align}
and are explicitly given by, 
\begin{align}
\ket{z}=D(z)\ket{0}=e^{za^\dagger-\bar{z}a}\ket{0}=e^{|z|^2/2}\sum\limits_{k=0}^\infty \frac{z^k}{\sqrt[]{k!}}\ket{k}.\label{eq:coherent}
\end{align}
where $z$ is a point in the (complexified) 2d phase space, and $D(z)$ are the displacement operators of the Heisenberg group that generate the coherent states\cite{Perelomov}. In the LLL language, using a radial gauge, $k$ is interpreted as the angular momentum. 

While the treatment of single-particle coherent states is standard, the two-particle coherent states are more subtle, particularly for the case of anyons\cite{Kjonsberg,Hansson}. 
The  two-particle coherent state associated with the coordinates $z_1$ and $z_2$ are products of single-particle states, $\ket{z_1,z_2}=\ket{z_1}\otimes\ket{z_2}$. 
 If the particles are distinguishable, this description is complete. But if the particles are indistinguishable the quantum statistics will effect also the classical theory. For instance, we will see that the Pauli exclusion principle has a classical counterpart. 
 
Since the quantum statistics only affects relative motion, it is convenient to transform to relative and center-of-mass (CM) coordinates. Using the 1d particle language, the canonical transformation from the original positions and momenta $(\vec{r_1}, \vec{p_1})$ and $(\vec{r_2}, \vec{p_2})$ to the CM and relative coordinates is given by
\begin{align}
\vec{R}&=\frac{1}{\sqrt{2}}(\vec{r}_1+\vec{r}_2)\quad ,\quad \vec{P}=\frac{1}{\sqrt{2}}(\vec{p}_1+\vec{p}_2)\\
\vec{r}&=\frac{1}{\sqrt{2}}(\vec{r}_1-\vec{r}_2) \quad , \quad\vec{p}=\frac{1}{\sqrt{2}}(\vec{p}_1-\vec{p}_2).
\end{align} 
Note that the CM and relative coordinates are presented here with modified normalization factors as compared to the usual convention (as in Ref.\cite{hill}). This modification does not change the formalism in any significant way, and will be used as notation in what follows.

There are two sets of ladder operators $a_c$ and $a_r$ each acting on the Hilbert space referring to the CM and relative motion respectively, and a total state is a tensor product between coherent states in each space as, \ie
\begin{align}
\ket{Z,z}=\ket{Z}_C\otimes\ket{z}_r
\end{align}
where $\alpha_c$ and $\alpha_r$ are complex parameters denoting the positions of the CM and relative coherent state respectively. The CM coherent state is just a regular one-particle state, while the relative coordinate coherent state encodes particle statics via the statistical boundary condition. In the simplest cases of bosons and fermions, we have
\be{bosferm}
\ket{z}_b= N_b (\ket{z}+\ket{-z})\quad , \quad
\ket{z}_f=N_f (\ket{z}-\ket{-z})
\ee
where $N_b$ and $N_f$ are normalization factors given below.
As an aside, the relative states in Equation \eqref{bosferm} are referred to as ``cat" states in the quantum optics literature \cite{DODONOV1974597}.
In the LLL language, the bosonic states are superpositions of even angular momentum states, while fermionic states are superpositions of odd angular momentum states in Equation \eqref{eq:coherent}. 

There is an immediate  generalization to an arbitrary $N$-particle bosonic or fermionic coherent state,
\be{ncoh}
    \ket{\{z_i\},\pm}=N(z_i,\bar{z}_i)\frac{1}{\sqrt{N!}}\sum_P\eta_P^\pm e^{z_{i_P}a_i^\dagger}\ket{0} \, ,
\ee
where the sum is over all  permutations with $\eta_P^+ =1$ for bosons and $\eta_P^- =\pm 1$, depending on whether the permutation is even or odd, for fermions. With this, th normalization becomes 
\begin{align}
    |N(z_i,\bar{z}_i)|^{-2}=\sum_P\eta_Pe^{\bar{z}_{i_P}z_i}, 
\end{align}
and the classical Lagrangian in holomorphic coordinates reads
\be{classlag}
   L(z_i,\bar{z}_i) =A_{\bar{z}_i}\dot{\bar{z}}_i+A_{z_i}\dot{z}_i-V(z_i,\bar{z}_i)
\ee
where $ V(z_i,\bar{z}_i) = \bra{\{z_i\},\pm}\hat{H}\ket{\{z_i\},\pm}$.

As pointed out in \oncite{hill}, if we consider the Hamiltonian $\hat H$ for the harmonic potential, $(\hat x^2 + \hat p^2)/2$, or $(\hat X^2 + \hat Y^2)/2$, the equations of motion following from Equation \eqref{classlag} are simply $\dot{z_i} = -i\omega z_i$ with no dependence on the statistics. Below we explore the situation for more general  Hamiltonians where the dynamics indeed depends on the classical statistics. Our examples will be a particle moving in an inverted 1d harmonic potential, and particles in the LLL subjected to an elliptic potential.

 \begin{figure}
    \centering
    \includegraphics[width=0.65\textwidth]{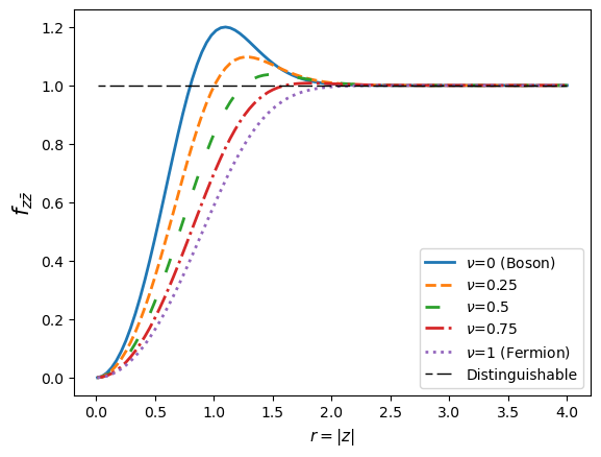}
    \caption{The symplectic factor $f_{\bar{z}z}=f(r^2)$ 
    defined in Equation \eqref{stwoform}  for anyons of exchange phase $\nu\pi$ as a function of the relative coordinates, in units of  $\sim \ell_B^{-2}$, which we have scaled to be dimensionless. This factor carries the crucial quantum statistics information going into the classical equations of motion. For distinguishable particles, the symplectic factor is always 1, which is also the asymptotic value for all identical particles. The difference between fermions and bosons is most pronounced for distances $\sim 0.5-2$ units.}
    \label{BF}
\end{figure}

Returning to two-particle systems, and defining  $Z=\frac{1}{\sqrt{2}}(z_1+z_2)$ and  $z=\frac{1}{\sqrt{2}}(z_1-z_2)$ so that $\ket{z_1}\otimes\ket{z_2}=\ket{Z}\otimes\ket{z}$, we have
\begin{align}
    \mathbb{K}_{CM}(Z,\bar{Z})&=Z\bar{Z}\\
    \mathbb{K}_b(z,\bar{z})&=\ln\cosh{z\bar{z}}\\
    \mathbb{K}_f(z,\bar{z})&=\ln\sinh{z\bar{z}}.
\end{align}
and from this,
\begin{align}
    f_{\bar{z}z}&=\tanh{z\bar{z}}+z\bar{z}\, \sech^2{z\bar{z}} \quad\quad \textnormal{ (Bosons)}\label{fb}\\
    &=\coth{z\bar{z}}-z\bar{z}\, \csch^2{z\bar{z}} \textnormal{ \quad (Fermions)}\label{ff}\\
    &=1 \hskip 4.3cm \textnormal{ (Distinguishable)}.\label{fd}
\end{align}
These symplectic factors are shown in Figure \ref{BF}. Plugging them back into the equations of motion in Equation \eqref{eom}, gives  a classical description of the dynamics which retain information about the quantum statistics.


\section{Scattering against an inverted oscillator in 1D} \label{1dho}

As discussed in \cite{hill}, to mimic the full QM system, the potentials involved should be weak. In the LLL interpretation this means that gradients are small compared to the magnetic length. In the 1d harmonic oscillator picture it would involve adding a small term,  as  $\sim \hat x \hat p$, to the Hamiltonian. We can, however, take another viewpoint and consider the classical system in Equation \eqref{classlag} in its own right and study how the dynamics depend on the statistics. In this case we simply ignore the connection to the original QM system and consider an arbitrary Hamiltonian. In this section we shall use  this approach, and exemplify with an inverted harmonic oscillator  potential since it will give an intuitive, and easily interpreted, picture of the effect of the statistics. In the next section we shall deal with the much more realistic case of particle moving in the LLL level while  subjected to a smooth external potential.

\begin{figure}
    \centering
    \includegraphics[width=\textwidth]{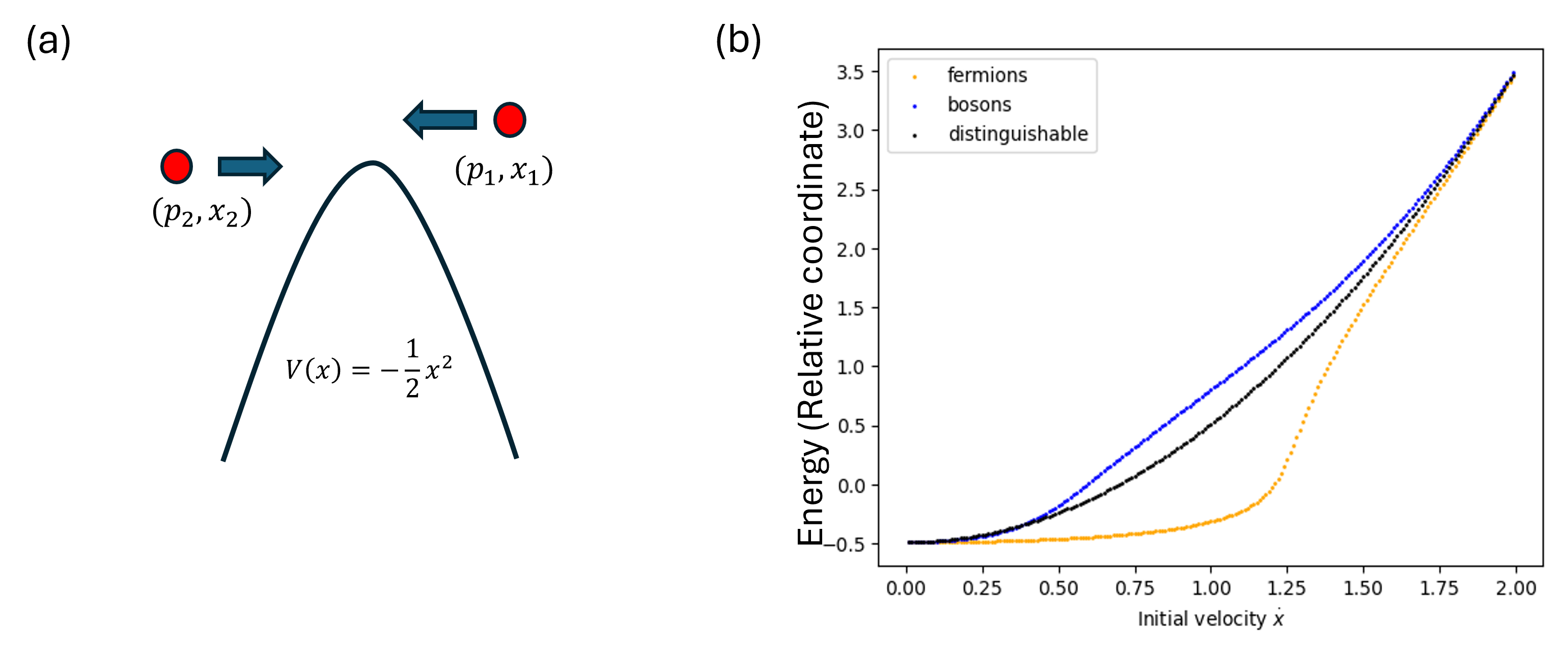}
    \caption{Scattering against a 1D inverted harmonic potential with $m\omega=1$: (a) A pictorial representation of two-particle scattering. The maximum of the potential is at $(x,p)=(0,0)$. For all trajectories, the center of mass is assumed to be located at $(X_{CM},P_{CM})=(0,0)$;  all dynamics is in the relative coordinate.  (b) Energy of each particle type for a fixed $x(0)$ and varying $\dot{x}(0)$  as described by Equation \eqref{ene}. It is seen that for a given initial position, the energies of the particles are ordered by type as $E_{bos}>E_{dist}>E_{fer}$, as expected from the magnitude of their symplectic forms in this regime.}
    \label{fig:cartoon}
\end{figure}

We shall use the dimensionless complex coordinate defined in Equation \eqref{complex1d}, and study two-body scattering in the inverse harmonic potential shown in Figure \ref{fig:cartoon}(a), \ie we use the negative sign in Equation \eqref{hamilton}. 

We have numerically integrated the previously given equation of motion,
\be{eom2}
    \dot{z}=-i\frac{\bar{z}}{f_{z\bar{z}}}
\ee
in the relative coordinate, 
\begin{figure}
    \centering
    \includegraphics[width=\textwidth]{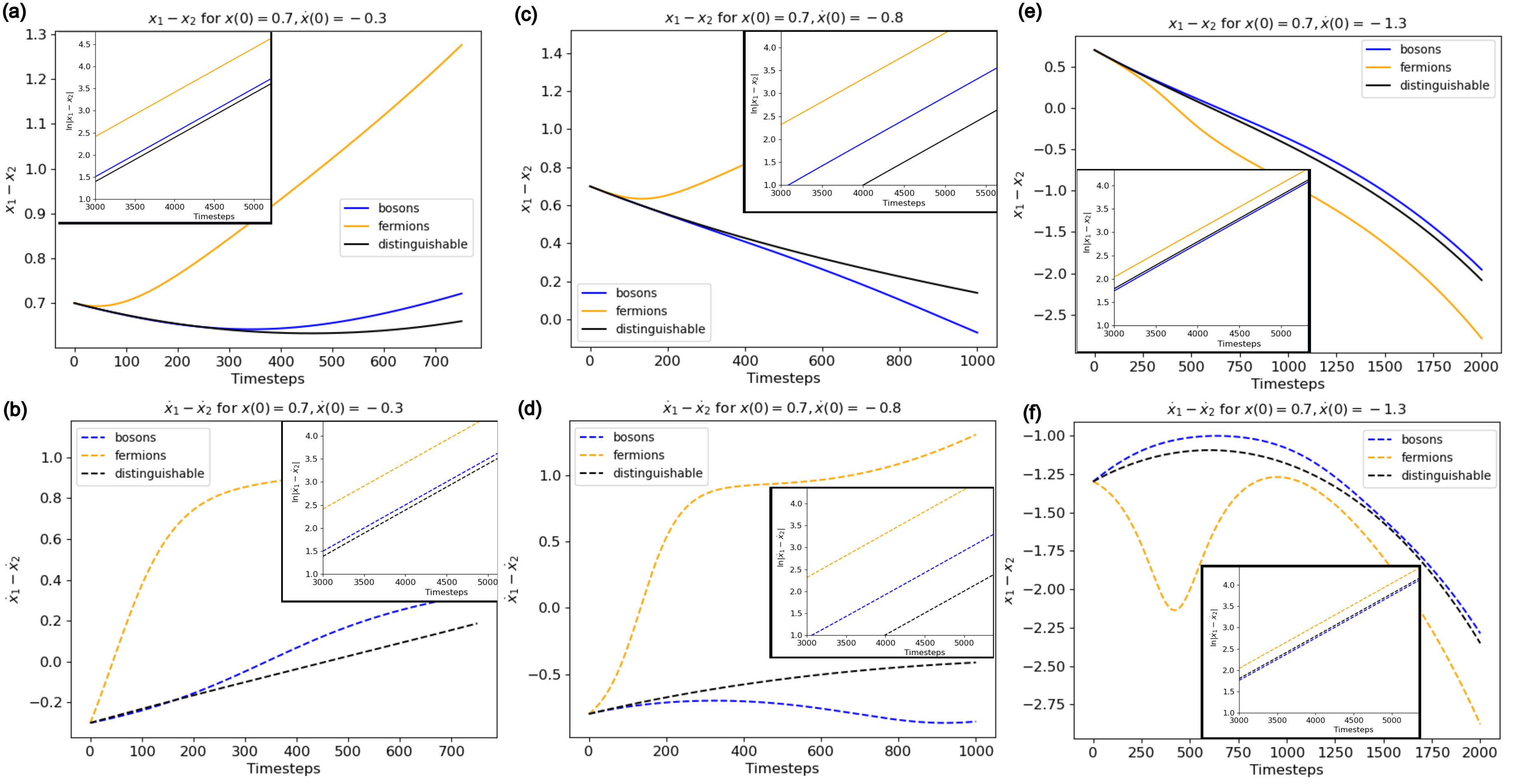}
    \caption{Scattering trajectories for two particles in the presence of an inverted harmonic potential with initial condition $x(0)=0.7$ and varying initial velocities.
   Time  is measured in dimensionless units, and the the scale of the x-axis is in 0.01 of this unit. The figures show the evolution of relative distance $x_1-x_2$ and relative velocity $\dot{x}_1-\dot{x}_2$ between the two particles respectively for initial relative velocity (a), (b) $\dot{x}(0)=-0.3$, (c), (d) $\dot{x}(0)=-0.8$ and (e),(f) $\dot{x}(0)=-1.3$. Insets: lin-log plots for large times. As initial velocities are increased, all particle species transition from reflecting against the potential to passing through each other. The critical initial velocity for this transition depends on the type of particle species under consideration.
    }
    \label{fig:IHO1}
\end{figure}
for three illustrative initial conditions, always with the center of mass fixed at $X_{CM} =  P_{CM} = 0$.

The resulting trajectories are shown in Figure \ref{fig:IHO1}.  For all three sets of initial conditions, the initial position is kept constant at $x(0)=0.7$, while the initial velocity $\dot{x}(0)$ is varied from $-0.3$ to $-1.3$. The trajectories vary significantly depending on the statistics of the particles, even when the initial conditions are the same. We can identify two qualitatively distinct behaviours: (i): The sign of the position trajectory remains the same throughout (positive in this case) while the velocity flips sign at a turning point, which is the point of closest approach.  (ii): The position changes sign while the sign of the velocity trajectory remains the same throughout (negative in this case). This corresponds to the particles simply passing through each other. Note that the fermionic repulsion is in {\it phase space} so nothing prevents the fermions to be at the same position as long as their velocities are different.

From Figure \ref{fig:IHO1}(a,b), we see that for small initial velocities, all particle species are scattered off the potential. Note that the distance of closest approach is larger for fermions than for bosons or distinguishable particles, which is a signature of their exclusion statistics. For higher velocities as in Figure \ref{fig:IHO1}(c,d), the fermions continue to scatter off the potential while bosons and distinguishable particles pass through each other. For even higher velocities, as in Figure \ref{fig:IHO1}(e,f), all species pass through each other. The markedly different behaviour of the fermion in the second set of initial conditions, as well as the trend of behaviours across all the initial conditions, can be qualitatively understood by considering the classical energy of each particle type for a given set of initial conditions:
\be{energy}\label{ene}
E=\frac{1}{2}(p^2-x^2)=\frac{1}{2}(f_{z\bar{z}}^2\dot{x}^2-x^2)
\ee
When the sign of the energy changes, this is equivalent to a flip in the qualitative nature of the  $x$ and $p$ trajectories, or in other words, $x$ and $\dot{x}$. From Equation  \eqref{ene}, it follows that the energy of a  particle  increases with velocity $\dot{x}$. That is, as we increase velocity through the three sets of initial conditions, the energies of the particles increase monotonically, as seen from Figure \ref{fig:IHO1}(b). When the energy changes sign from negative to positive, the nature of the position and velocity trajectories change as well. That is, for negative energy, particles scatter off the potential (case(i)) and for positive energy, particles pass through each other (case(ii)). However, as seen in Figure \ref{fig:IHO1}(b), fermions always have lower energy than bosons for a given set of initial conditions since the symplectic factor $f_{z\bar{z}}$ for fermions is always smaller than that of bosons (See Figure \ref{BF}). This energy ordering implies that the change in sign of the energy for fermions happens at a much higher initial velocity than for bosons, which explains the trajectories in Figure \ref{fig:IHO1}(c,d). 
 
 It is also notable that for the final set of initial conditions with the largest energy, all trajectories are much closer to that of distinguishable particles than for the lower energies. This feature is again explained by considering the nature of the symplectic factor. As seen in Figure \ref{BF}, the symplectic factor for both fermions and bosons approaches 1 for large relative distance/velocity between the two particles. That is, their behavior approaches that of distinguishable particles.

 The suppression of features arising from the symplectic factors is even more pronounced at large times (and thus, large separations) when the equations of motion for all three particle types are identical. The classical solutions are easily obtained as $x(t)\sim\dot{x}(t) \sim e^t$. 
 
 This feature is represented as the inset lin-log plots in Figures \ref{fig:IHO1}(c-e). For large times, all particle types, for all initial conditions, have linear trajectories with identical slopes in the log-log plots, thus demonstrating the suppression of all statistics effects. 

 To summarize, the quantum statistics of the different particle types is encoded in their respective symplectic factors, and thus in their classical equations of motion, and affect the trajectories in a way that can be qualitatively understood. In other words, the effective bunching or exclusions  characteristics, of these particles are reflected in the classical trajectories that are obtained from the equations of motion.

\begin{figure}
    \centering
    \includegraphics[width=0.3\textwidth]{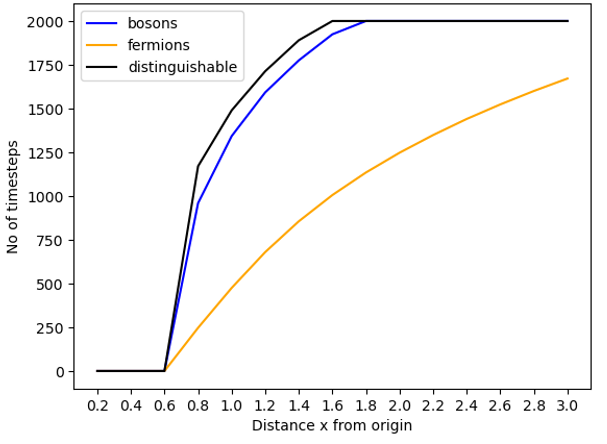}
    \includegraphics[width=0.3\textwidth]{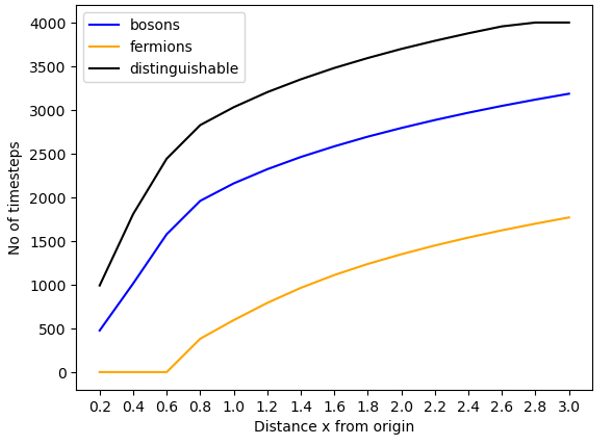}
    \includegraphics[width=0.3\textwidth]{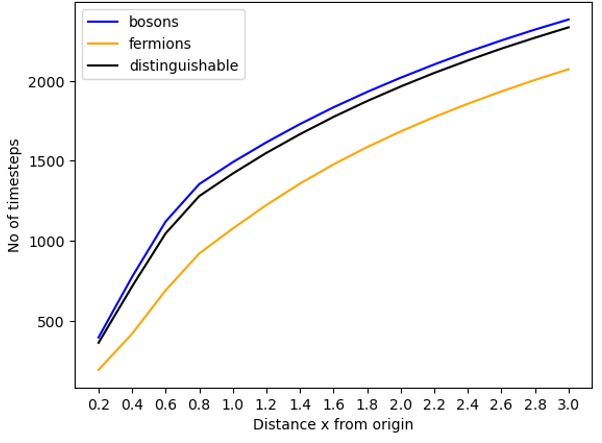}
    \caption{The  function $R(x_L)$ defined as the number of time steps for which the trajectory $x(t)<x_L$ for the initial conditions: $x(0)=0.7$ and $\dot{x}(0)=-0.3$, $x(0)=0.7$ and $\dot{x}(0)=-0.8$ and $x(0)=0.7$ and $\dot{x}(0)=-1.3$, for the left center and right panels respectively. Note that the fermions spend less time in any given interval than the other particle types.}
    \label{fig:IHO2}
\end{figure}
 
 We further emphasize the importance of the ``classical statistics" by defining a ``survival parameter" for each set of trajectories. Inspired by measures of survival probability in decay problems, we define the function $R(x_L)$ as the amount of time $t$ during which $x(t) < x_L$. The $R(x_L)$ function for each set of initial conditions considered above is given in Figures \ref{fig:IHO2}, and  shows that for all the  initial conditions, the fermions always spend less time in any interval than do the bosons. This illustrates the higher ``repulsive” or ``anti-bunching” tendency of fermions, as compared to bosons, since they spend comparatively shorter time close to each other. 


\section{Lowest Landau level Coherent states in quadratic potentials} \label{lllcs}

We have stressed in previous sectionsthat guiding center motion of charged particles in a strong magnetic field, with an obvious connection to the quantum Hall effect, is the prime physical system where our formalism can be applied. In this section we study the trajectories of classical fermions and bosons in harmonic potentials and contrast their behaviour with that of distinguishable particles. 

We start with a brief description of the physics in the LLL, beginning with the Hamiltonian, 
\begin{align}
    H_{2DEG}&=\sum_i \frac{(\vec{p}_i-e\vec{A}_i)^2}{2m}+V_{pot}= \sum_i \frac{\vec{\Pi}_i^2}{2m}+V_{pot}\\
    &=\sum_i\hbar\omega_c\big(\hat{n}_i+\frac{1}{2}\big)+V_{pot}
\end{align}
where the cyclotron frequency $\omega_c =B/m$ sets the energy scale. The LLL, which amounts to $n=0$, is infinitely degenerate, and, in a radial gauge, the  one particle states are labelled by the angular momentum quantum number. It is convenient to introduce the guiding center coordinate operators, 
\be{guide}
\hat X = \half \hat x +\frac 1 B \hat p_y \quad , \quad \hat Y = \half  y - \frac 1 B \hat p_x
\ee
satisfying the canonical commutation relations,
\begin{align}
    [\hat{X},\hat{Y}]&=-i\ell_B^2.
\end{align} \ 
The guiding center coordinates therefore constitute a non-commutative plane. The corresponding  
ladder operators $a$ and $a^\dagger$, where 
\begin{align}
    a=\frac{1}{\ell_B}(\hat{X}+i\hat{Y}) \, ,
\end{align}
lower and raise the angular momentum quantum number.  As defined in Equations \eqref{eq:cs}-\eqref{eq:coherent} of Section \ref{CS}, coherent states of this algebra correspond to the specific superposition, $\ket{z}=e^{|z|^2/2}\sum\limits_{k=0}^\infty \frac{z^k}{\sqrt[]{k!}}\ket{k}$ of the angular momentum states.

Since the kinetic energy is quenched  in the bulk of the LLL,  the projected form of the applied potential fully determines the effective Hamiltonian in the. We will assume that the strength of the potential is much smaller than the cyclotron frequency, ensuring that the system remains in the LLL under time evolution. Therefore, given a potential $V_{pot}=V(\hat x,\hat y)$, we can determine the effective Hamiltonian to be
\begin{align}
    H_{LLL}&=P_{LLL}V(\hat x,\hat y)P_{LLL}=V(\hat{X},\hat{Y})\\
    &=V(a,a^\dagger)
\end{align}
where $P_{LLL}$ are projection operators onto the lowest Landau level. 

We consider a generic quadratic potential of the form 
\begin{align}
    V&=\frac{1}{2}u(x_1^2+x_2^2)+\frac{1}{2}v(y_1^2+y_2^2)\\
    &=\sum_{i=1,2}\frac{U}{4}(\bar{z}_iz_i+z_i\bar{z}_i)+\frac{V}{4}(z_i^2+\bar{z}_i^2)\\
    &=\frac{U}{4}(\bar{Z}Z+Z\bar{Z})+\frac{V}{4}(Z^2+\bar{Z}^2)+\frac{U}{4}(\bar{z}z+z\bar{z})+\frac{V}{4}(z^2+\bar{z}^2) , 
\end{align}
 where we used the notation of Section \ref{background}. Also $P=\ell_B^2(u+v)$ and $V=\ell_B^2(u-v)$. 

Using the coherent state formalism described in Section \ref{CS}, and taking $z_i, z, Z$ to be the eigenvalues of the annihilation operators $a_i, a, A$ respectively, the quadratic potential operator can be written, 
\begin{align}
    V&=\frac{U}{2}(A^\dagger A+\frac{1}{2})+\frac{V}{4}(A^2+{A^\dagger}^2)+\frac{U}{2}(a^\dagger a+\frac{1}{2})+\frac{V}{4}(a^2+{a^\dagger}^2) \, .\label{Ham}
\end{align}
When $u=v>0$, the potential is a symmetric harmonic trap, and when $u\ne v$ and $u,v>0$, it is an asymmetric (or elliptical) harmonic trap. When $u$ and $v$ have opposite signs, we have a saddle potential, which again is symmetric for $u=-v$.  The analysis below holds for all these cases.

To summarize, we give the Lagrangians for the $CM$ coordinate and the relative coordinate for the classical bosons and fermions respectively, using Equation \eqref{bosferm}.
\begin{align}
    L_{CM}&=\frac{i}{2}(\dot{Z}\bar{Z}-\dot{\bar{Z}}Z)-\frac{U}{2}Z\bar{Z}-\frac{V}{4}(Z^2+\bar{Z}^2)\\
    L_b&=\frac{i}{2}\tanh{z\bar{z}}(\dot{z}\bar{z}-\dot{\bar{z}}z)-\frac{U}{2}z\bar{z}\tanh{z\bar{z}}-\frac{V}{4}(z^2+\bar{z}^2)\\
    L_f&=\frac{i}{2}\coth{z\bar{z}}(\dot{z}\bar{z}-\dot{\bar{z}}z)-\frac{U}{2}z\bar{z}\coth{z\bar{z}}-\frac{V}{4}(z^2+\bar{z}^2).
\end{align}

\subsection{Phase space trajectories}
The corresponding equations of motion are given by
\be{cmeom}
    \textnormal{CM: }\dot{Z}&=-\frac{i}{2}\big(UZ+V\bar{Z}\big)\\
    \textnormal{Relative: }\dot{z}&=-\frac{i}{2}\big(Uz+V\frac{\bar{z}}{f_{\bar{z}z}}\big)\label{releom}
\ee
where the symplectic factor $f_{\bar{z}z}$ is given by Equations \eqref{fb}--\eqref{fd}.

Recall from Figure \ref{BF}, that for long inter-particle distances, the symplectic factor for both fermions and bosons is identical to that for distinguishable particles, and, as is intuitively clear, we do not have any statistics dependence.   More surprisingly, both factors approach zero for small distances where one might have expected the largest deviation. 

The length scale where the differences between bosons and fermions becomes most apparent is in the 0.5-2 units range; we show trajectories in this range to highlight these differences. 
\begin{figure}
    \centering
    \includegraphics[width=0.45\textwidth]{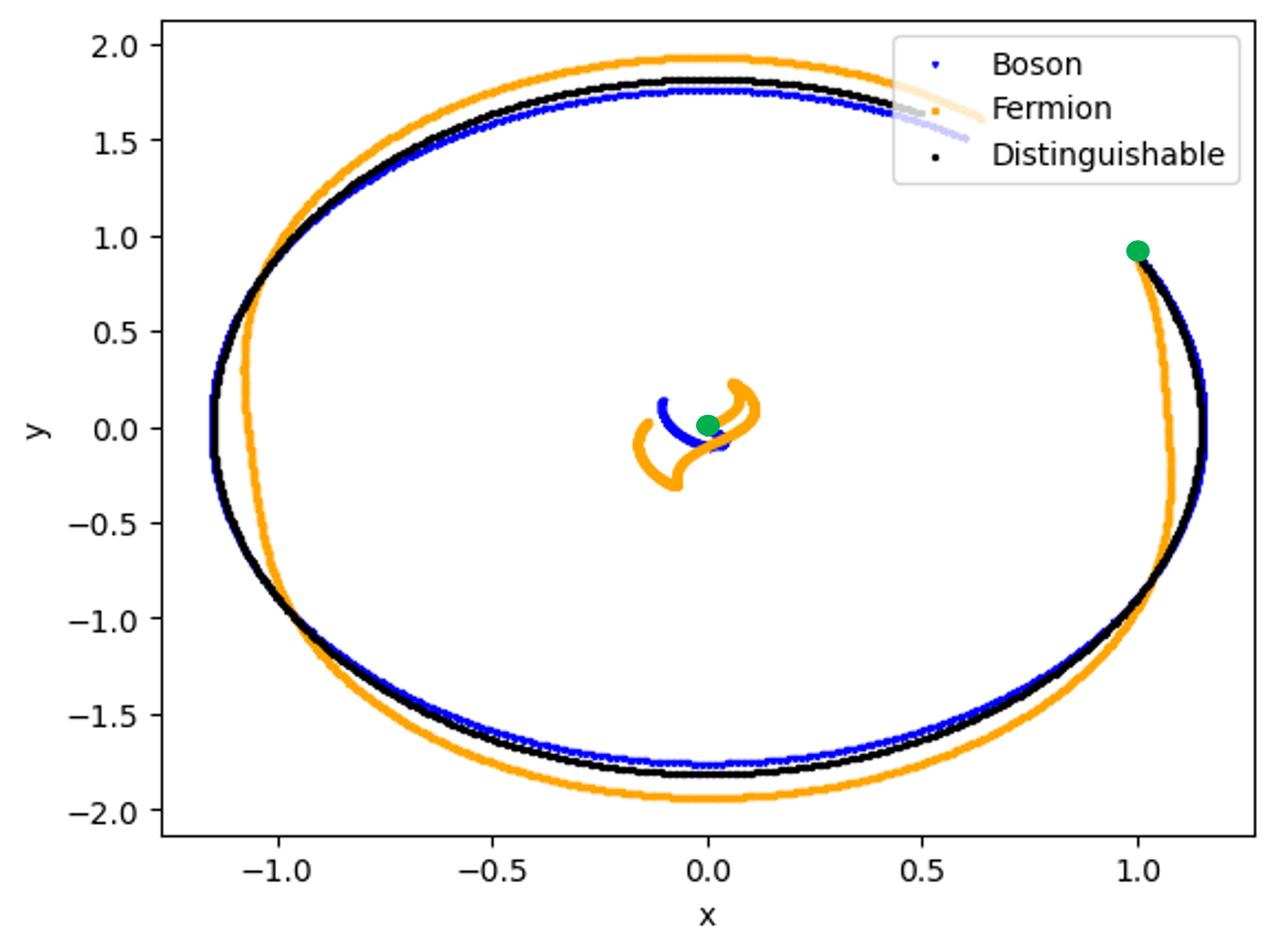}
    \includegraphics[width=0.45\textwidth]{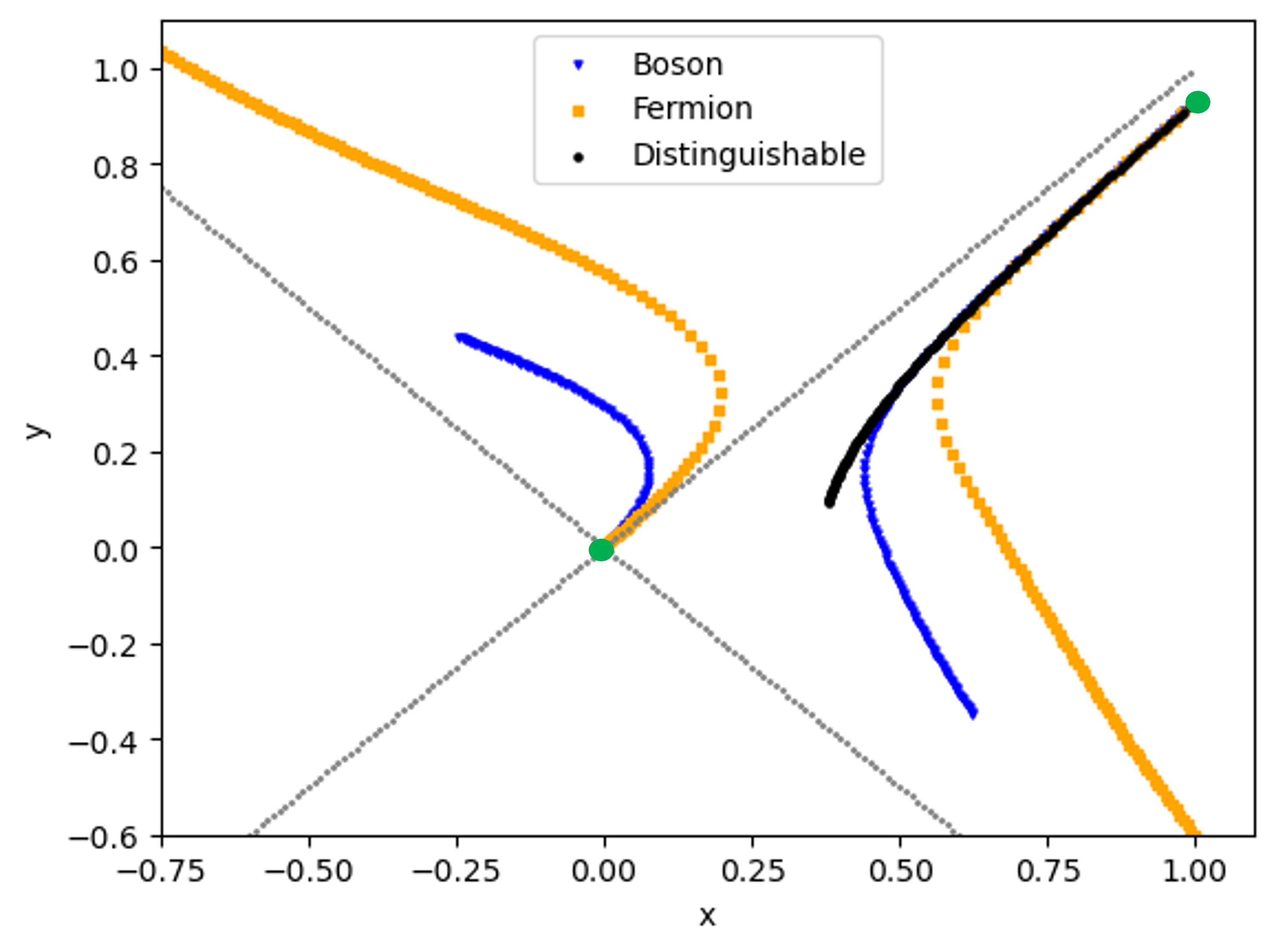}
    \caption{Trajectories for harmonic elliptic  and saddle potentials. In each case, the initial conditions for the two particles are shown as an enlarged green dot. Left Panel: All particles - single, distinguishable and identical - behave in a similar way when subjected to an elliptic harmonic trap with $v/u=0.4$. one particle is initially at the origin and the other at a distance $\sim$ 1.4 units away. Right Panel: A symmetric saddle potential with $u=-v=1$, again with one particle starting at the origin.  The grey lines $y=\pm x$ are the asymptotes of the applied saddle potential.
    }
    \label{elliptic}
\end{figure}
We have numerically integrated Equations \eqref{cmeom} and \eqref{releom} to obtain these trajectories, with the initial value $x=y=0$ for one of the particles, and with the other one placed at a distance $\sim 1-2$ units from the origin. As can be seen from Equation \eqref{releom}, when $u=v$ ($V=0$), the symplectic factor does not enter the equations of motion. In this symmetric situation, for all cases, one particle remains at the origin while the other moves in a circle around the origin. 

Figure \ref{elliptic} shows the motion of the particles in an elliptical potential with $v/u=0.4$, as well as the motion in a symmetric saddle potential with $u=-v=1$ for the same initial conditions. For distinguishable particles, the particle at the origin remains at rest while the other moves along equipotential lines. On the other hand, both bosons and fermions behave quite differently. The particles at the origin in both cases do not remain at the origin anymore. They get displaced due to an effective ``statistical interaction" (we choose not to use the term ``force" since it implies encoding in the Hamiltonian). In general, for the same time interval, the fermions moves much further away from each other than the bosons, again due to this statistical interaction as seen in Figure \ref{elliptic}(b). 

\begin{figure}
    \centering
    \includegraphics[width=0.4\textwidth]{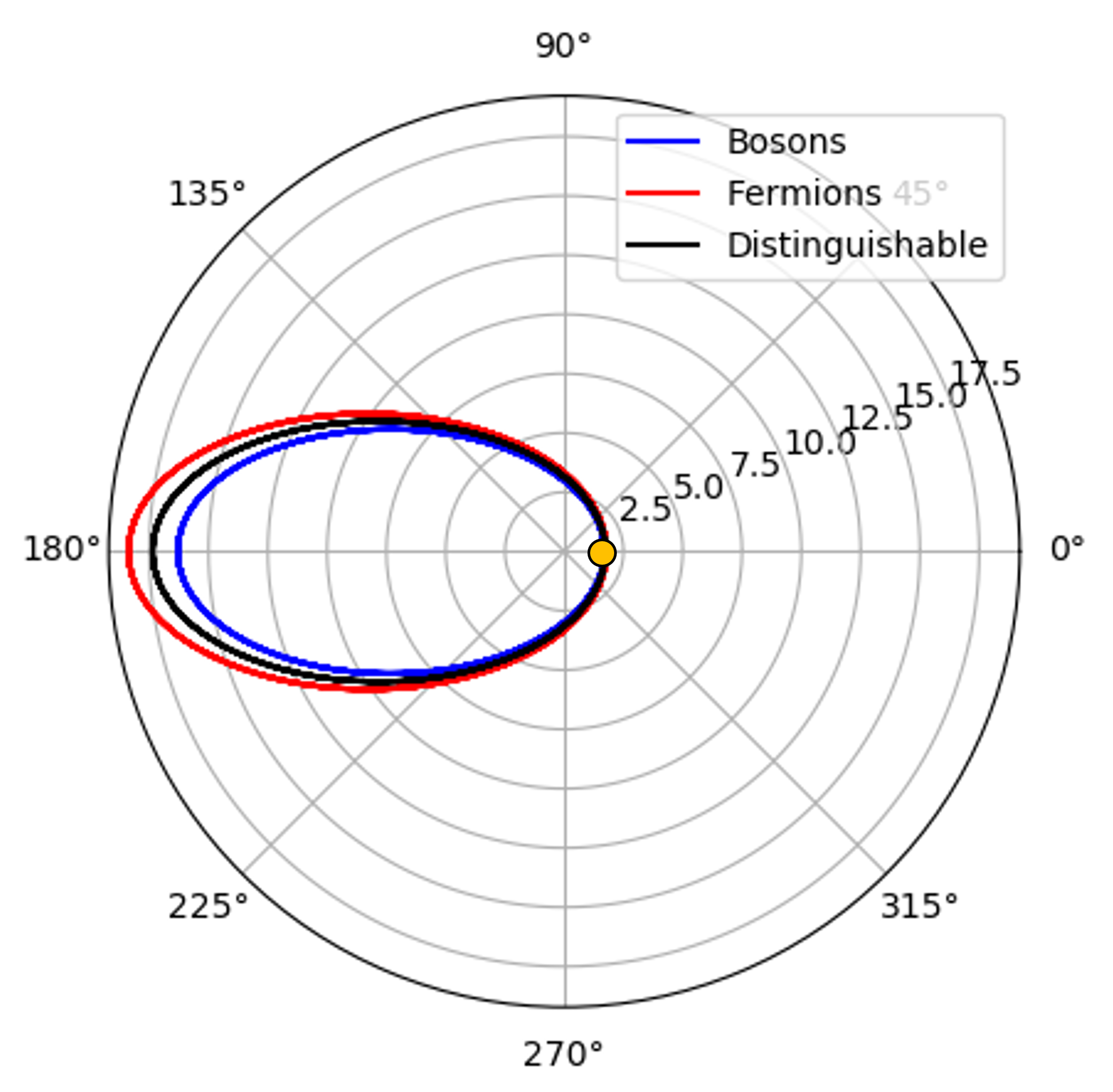}
    \includegraphics[width=0.5\textwidth]{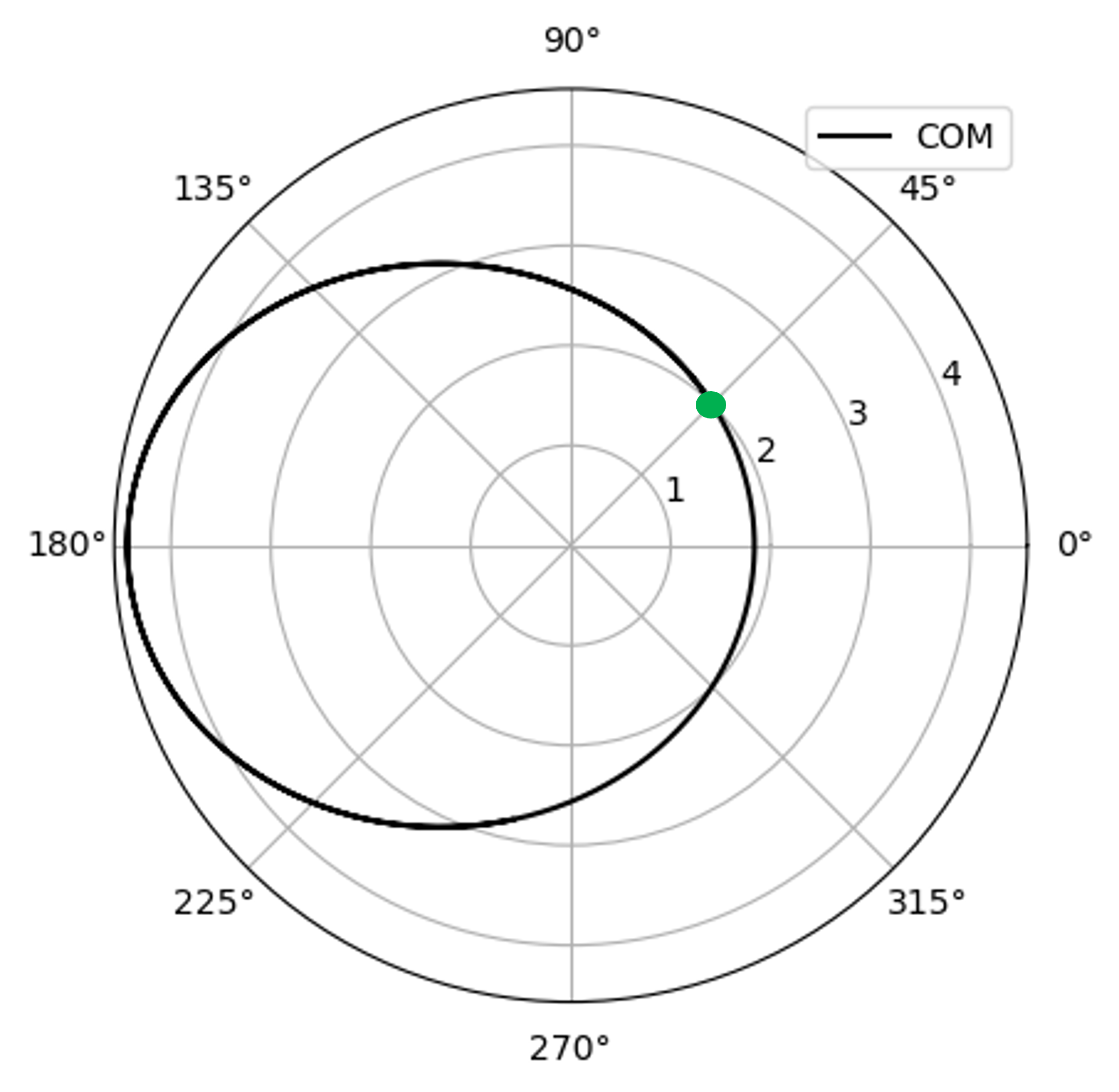}
    \caption{Left Panel: Relative coordinates trajectories, $r(\theta)$, for different statistics particles mowing in an elliptical trap are shown as a polar plot in $\rho=r^2$ and $\phi=2\theta$ coordinates. The equipotential surfaces of bosons encloses a smaller volume than fermions for the same initial conditions $(x,p)$. Initial position is indicated as an orange dot. Right Panel: The trajectory of the CM coordinate which is independent of statistics. The initial position is shown as a green dot.}
    \
    
    \label{ellipse1}
\end{figure}

This kind of behaviour is because the velocities $\dot{z}$ depends on statistics, as seen in Equation \eqref{releom}. Thus, at a given time, the relative coordinate of bosons, fermions and distinguishable particles can be vastly different, as is most clearly seen in the case of the elliptical trap. Instead of integrating the equations of motion  in time, one can derive an equation for the trajectories $r(\theta)$ and $R(\theta)$, which are shown  in Figure \ref{ellipse1}.

\begin{figure}
    \centering
    \includegraphics[width=\textwidth]{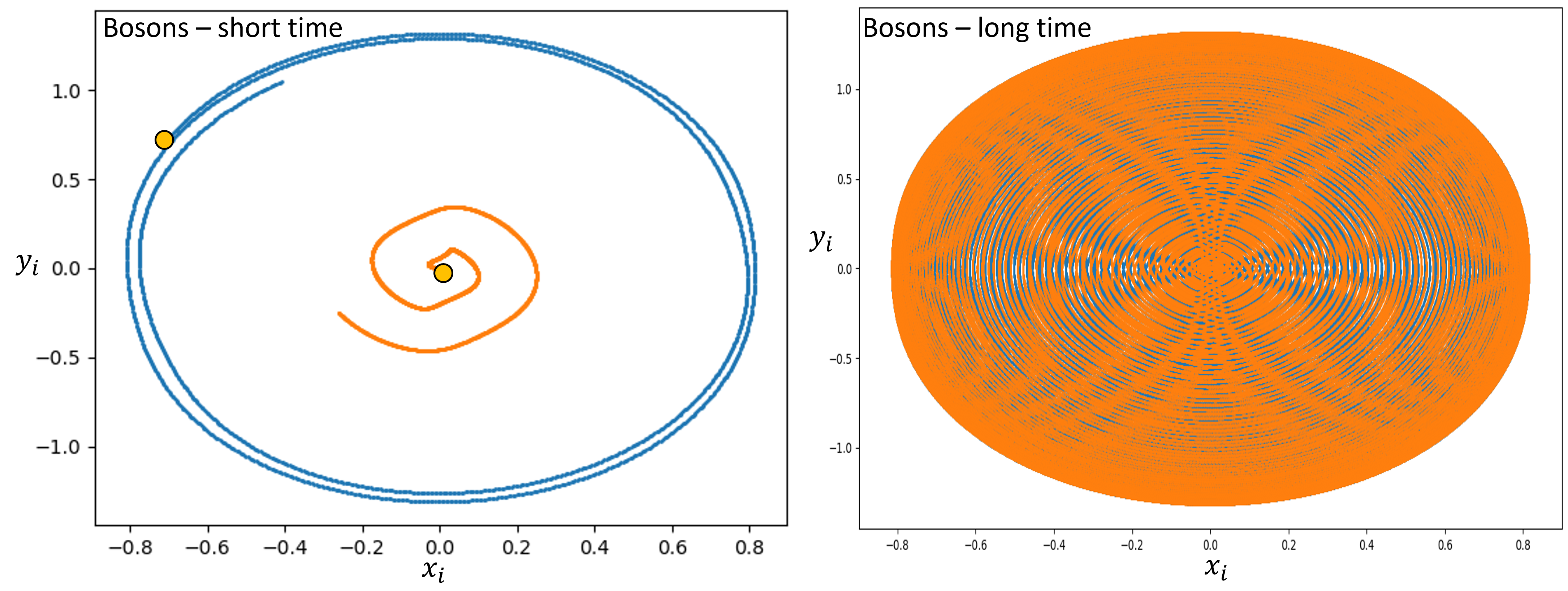}
    \includegraphics[width=\textwidth]{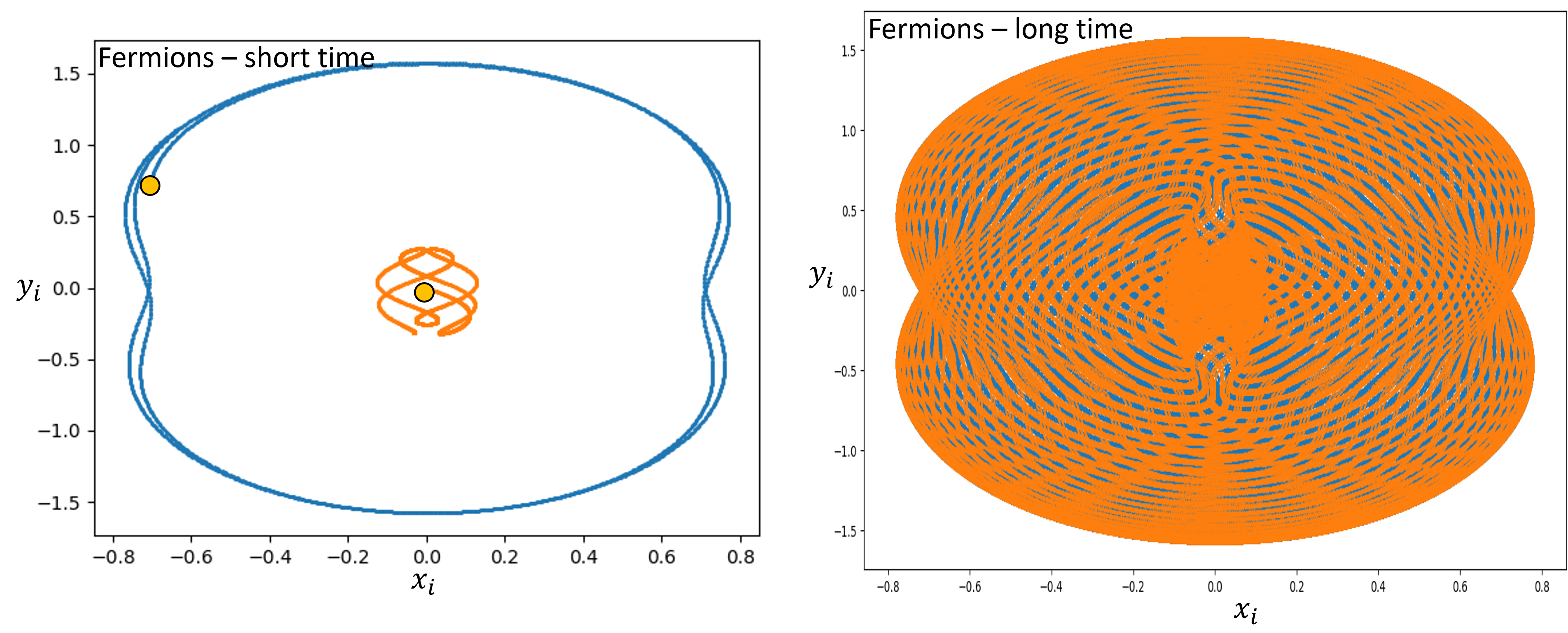}
    \caption{Identical particles in an elliptical trap with $v/u=0.4$ with initial positions marked by yellow dots.  Top panels: For bosons, the particles spiral back and forth towards each other. After a very long time, the trajectories densely cover the region where their initial total energy is conserved. Bottom Panels: For fermions, the particles also spiral back and forth towards each other, although not obvious from the figure, by studying the time evolution we  have seen that fermions tend to avoid each other at close approach leading to a very different trajectory as compared to bosons. After a very long time, the trajectories again densely cover the region allowed by  energy conservation. The volume of the effective phase space covered by fermions at a given energy is larger than that of bosons.}
    \label{ellipseA}
\end{figure}

As expected, all curves are closed, bounded ellipses. However, as we discussed above, the velocities depend on statistics, and hence the periods of the orbit. The effects of this dependence are clearly seen when we map back the trajectories from the $z$ and $Z$ coordinate to the individual particle coordinates $z_1$ and $z_2$. When the center of mass is at rest \ie $Z(t=0)=0$, it has no dynamics and remains at rest. In this case, the individual particle trajectories are also closed, bounded ellipses. However, if the center of mass is not at rest, the resultant individual particle trajectories are vastly different. In Figure \ref{ellipseA} we show the short and long time behaviour or the orbits for bosons and fermions.

Neither for  bosons nor fermions, do the trajectories close. Instead the particles spiral towards each other, albeit in very different ways. In particular, simulations of the trajectories observed in real time show that fermions tend to avoid each other on close approach and appear to bounce away from each other. Videos of such simulations are included as supplementary material. After a very long time, and in both cases, the trajectories tend to fill the region of the plane that is  allowed by energy conservation. 
We shall return to the long time behaviour in Section \ref{chaos}, where we speculate about chaotic behaviour in a three particle system.


\section{The possibility of chaos in elliptical potentials} \label{chaos}

Classically, the motion of free particle in a quadratic potential is perfectly regular. But 
as we have seen in previous sections, the trajectories obtained by projecting particles with fermionic or bosonic exchange statistics results in non-linear equations of motion. Specifically in the case of anisotropic bounded potentials like the ellipse, the trajectories seem to be dense over the equipotential surfaces. In this section we ask if these dense trajectories are actually an example of deterministic chaos.

Let us recall, the Liouville-Arnold theorem on integrability, which heuristically states that a system is fully integrable if the number of independently conserved integrals of motion $I$ equals the degrees of freedom of the system $N$ \cite{Zaslavsky}. If a (non-linear) system is not integrable, i.e., $I<N$ then it could have chaotic regimes.

The system consisting of two particles in the lowest Landau level has $N=2$. Though the individual energies of the particles are not conserved due to their exchange statistics, the total energy is conserved. Additionally, the energy of the center of mass is conserved, thus yielding two independently conserved quantities i.e., $I=2$. This makes the two-particle system fully integrable and not chaotic. 

However, in the case of three or more particles in the lowest Landau level, the degrees of freedom equals the number of particles $N\geq3$, but the number of independently conserved quantities remains $I=2$. The total energy of all particles and the energy of the center of mass are the only conserved quantities. Thus, these systems could exhibit deterministic chaos in certain regimes defined by their initial conditions. 

While a full identification and examination of chaotic regimes in the three (or higher) particle system is beyond the scope of this work, we here present here some preliminary numerical data that suggests that there is such chaotic regime in the three-particle case. Our analyses show that these particles that carry a memory of quantum statistics can access chaotic regimes under conditions in which purely classical particles would not be able to do so. Specifically, for our choice of potentials, the latter would exhibit closed orbits, unlike with scenarios involving billiard table- or stadium-type confinement. Furthermore, unlike with chaos in the standard three-body problem, the particles here are non-interacting. As we have seen before, the source of non-linearity in the corresponding equations of motion associated with this non-interacting system is the symplectic form, and it provides the source of chaos here.


Since trajectories are always fully integrable in a symmetric harmonic potential, we consider elliptical potentials in the LLL as in Section V. 
We integrate the equations of motion for an elliptical potential  at $v/u=0.4$ and two sets of initial conditions that differ  by $\sim1.2\%$. We estimate the Lyapunov exponent in the 6-dimensional phase space in which the trajectories are embedded using the standard expression\cite{ChaosBook} 
\begin{align}
    \lambda=\lim_{t\to\infty}\lim_{\Delta \tilde Z(0)\to 0}\frac{1}{t}\ln{\frac{\Delta \tilde Z(t)}{\Delta \tilde Z(0)}}
\end{align}
\begin{figure}
    \centering
    \includegraphics[width=\textwidth]{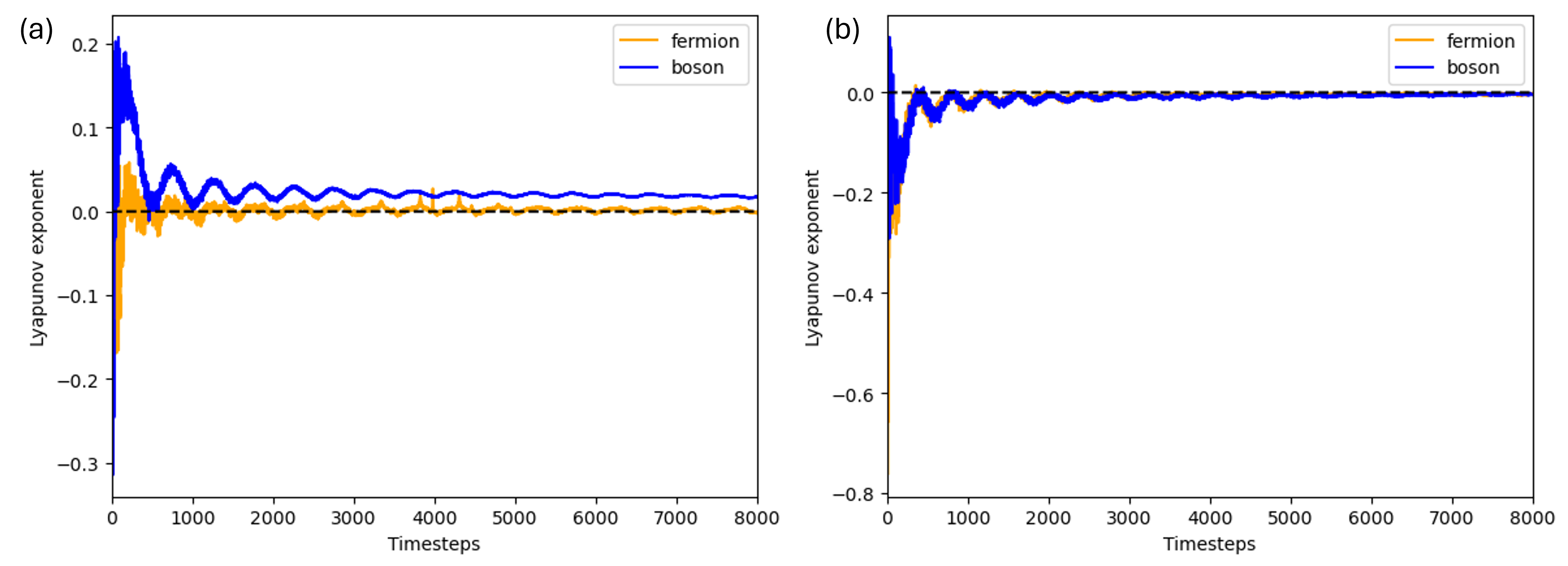}
    \caption{Estimates of the Lyapunov exponents for 3-particle trajectories in an elliptical potentials with  (a) $v/u=0.4$ and (b) $v/u=0.6$. In each case the particles are arranged symmetrically around the origin, and the initial conditions differ by $\sim 1.2\%$.}
    \label{ch1}
\end{figure}
where $\Delta \tilde Z(t)$ is the separation between the two trajectories in the six dimensional space at time $t$. From Figure \ref{ch1}(a), we see that at longer times, the bosonic trajectories appear to converge to a non-zero Lyapunov exponent  $\lambda\sim 0.02$ while fermions have $\lambda=0$, indicating that bosons exhibit deterministic chaos in this region. On repeating this process for an elliptical potential of lower eccentricity, as in Figure \ref{ch1}(b), we see that $\lambda$ disappears for both particle species, meaning that we are outside the chaotic regime also for bosons. 


This picture hints at how deterministic chaos does indeed seem to be present for greater number of particles in certain regimes, and already points to interesting open questions, such as bosons appearing more prone to chaos than fermions, as can be gleaned from the Lyapunov analysis above. Furthermore, 
a full analysis determining the chaotic regimes in terms of ellipse eccentricity as well as initial conditions for both particle species is needed. {While the non-zero Lyapunov exponent  for this  specific set of parameters  suggests the existence of classical chaotic regimes in three-body dynamics, a more complete analysis is needed to fully characterize and quantify them, e.g. by analyzing relevant Poincare sections. Such an analysis is beyond the scope of this paper, but is left for future work. }


  \section{Quantum versus Classical dynamics } \label{quantcomp}

{In contrast to the classical formalism presented here, direct quantum mechanical time evolution of the fermionic and bosonic states is governed by the unitary operator $U=e^{-iHt}$ acting on the coherent states and has been studied in earlier works\cite{b3,b4}. Generally speaking, the trajectories obtained in each case will quantitatively differ from each other. Even in the simple case of single particle coherent states, where, as one might expect, the equations of motion are identical for classical and quantum dynamics their solutions are still give distinct trajectories. A  comparison of this quantum mechanical evolution with the classical approach pursued in this work is given in the Appendix. However it must be noted that such differences between the classical and quantum dynamics are an inevitable feature of the states under consideration and do not prevent us from gaining insights through the comparison of the fermionic and bosonic manifolds as shown in earlier sections.

In this section, we put our work in perspective, by recapitulating some general results about the geometric interpretation of quantum mechanics and then discussing our results in this more general context.}

When studying the mechanics of a classical system, it is useful to identify the state of the system as a point
on a symplectic manifold representing the phase space. The symplectic structure of this manifold endows  it with a
Poisson bracket, and observables are smooth differentiable functions over the manifold. The set of observables
is equipped with a commutative and associative algebra. The observables generate flows  on the manifold, and a particular one, the Hamiltonian, generates time evolution and thus  dynamics. In this sense, classical mechanics can be viewed as a geometric theory.

Quantum mechanics, on the other hand, has states that are vectors in a Hilbert space. Observables in quantum
mechanics are Hermitian operators over the Hilbert space and the system has an associated  Lie algebra defined by the commutator bracket. Again, the Hamiltonian is an operator that generates time translation of the states, and this evolution is linear However, physical quantum states are part of the projected Hilbert space or the quantum phase space, which is non-linear, and is a K\"{a}hler manifold endowed with the Fubini-Study metric as defined earlier in this work.

There has been extensive literature establishing the geometric interpretation of quantum mechanics over the decades\cite{geom1,geom2,geom3,geom4}. Crucially, the non-commutative nature of quantum mechanics which sets it apart from classical mechanics is encoded in the geometry of the phase space through the symplectic form. Several authors have pointed out that up to a constant, the following map holds true in both finite and infinite dimensional systems
\be{map}
    \{F,G\}_\omega=\frac{1}{i\hbar}\langle [\hat{F},\hat{G}]\rangle
\ee
where $\hat{F}$ and $\hat{G}$ are hermitian operators that become functions F and G on the phase space generated by a set of states $\ket{\psi}$ as $F=\bra \psi \hat{F}\psi\rangle$. The symplectic form on this manifold is $\omega$. Crucially,  this exact map between the commutator and Poisson bracket on the phase space is an identity, not a quantization rule. 

These considerations are so far completely general and amount to an alternative formulation of quantum mechanics that does not involve a Hilbert space. What we studied in this paper is a similar but conceptually distinct construction. We have not attempted to describe the full Hilbert space, but only a submanifold chosen so that time evolution, to a good approximation, takes place within this submanifold.

We  considered the submanifolds defined by the symmetry properties of the wave functions, which is preserved under time evolution, and then specialized to fermionic, bosonic or anyonic multi particle coherent states in the LLL, which were shown in \oncite{hill}, to be K\"ahler.

In this case, however, the map in Equation \eqref{map} is not an identity. To see this, note that the bosonic and fermionic relative coordinate coherent states considered in this work are special in that the expectation values of linear operators such as $\braket{\hat{x}}\sim \braket{a+a^\dagger}$ are identically zero. This follows rather straightforwardly by evaluating the expectation values for the bosonic and fermionic states in Equation \eqref{bosferm}. Therefore, on picking $\hat{F}=a+a^\dagger$ and $\hat{G}=i(a-a^\dagger)$, the LHS of Equation \ref{map} is zero, whereas their commutator being a non-zero constant results in a non-zero RHS. 

\begin{figure}
    \centering
    \includegraphics[width=\textwidth]{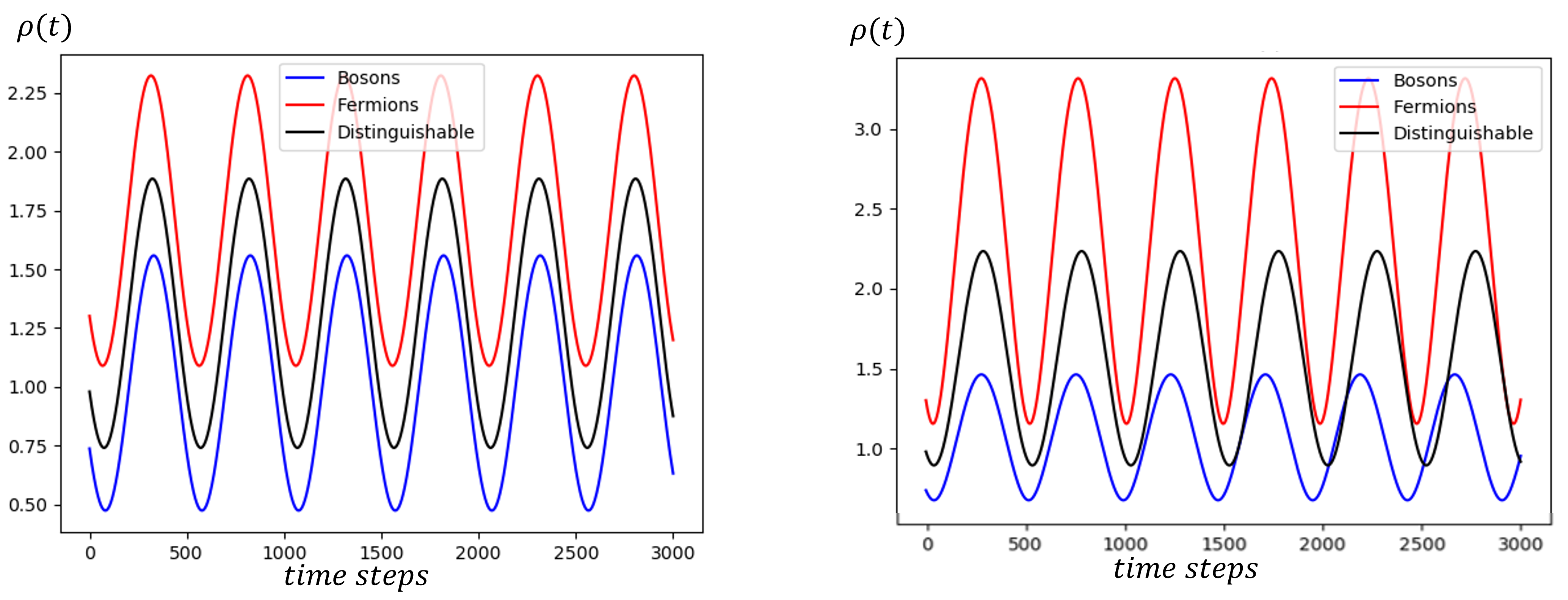}
    \caption{The coherent state expectation value and the effective phase space trajectory for identical particles are not equivalent. The function $\rho(t)=\langle  r^2(t)\rangle$ is plotted for fermions, bosons, and for reference, also for distinguishable particles, the coherent state approach (Left Panel) and the phase space method (Right Panel). Unlike in the case of distinguishable particles, where the mismatch was mere shift,  there are significant differences in terms of amplitude as well as phase. Moreover, all particle types have the same period in the coherent state approach, which is not the case for the phase space method. }
    \label{mismatch1}
\end{figure}

Therefore, instead of treating such classical dynamics over the phase space as an exact mapping of the  quantum mechanics, it is useful to view this evolution in terms of  accumulation of quantum phases. The relevant concept here is the Anandan-Aharanov phase\cite{geom3} which is topological rather than geometrical in nature. In our case, the total phase accumulated during a cycle  is, 
\begin{align}
    \phi=\int_0^T  dt\, \bigg(\bra{\psi}i\partial_t\ket{\psi}-\bra{\psi}\hat{H}\ket{\psi}\bigg)  \, ,
\end{align}
where the first term is the Anandan-Aharanov phase and the second is the dynamic phase. The total phase  is nothing but the classical action on the effective phase space!  Both of these terms would be different for the various particle species, but the Anandan-Aharanov phase,  which is evaluated on phase space, is independent of the Hamiltonian when the evolution is periodic and  can be rewritten as 
\begin{align}
    \int_0^T i\bra{\psi}\partial_t\ket{\psi}=\oint A_\xi.d\xi=\int f_{z\bar{z}}dzd\bar{z}
\end{align}
where $\xi$ are the coordinates of the phase space. This expression is explicitly different for the various types of statistics, and thur affects the  periodicity of the system. We demonstrate this  in Figure \ref{mismatch1} where the classical phase space trajectory corresponding to $\rho(t)=r^2(t)$ for distinguishable particles has the same periodicity as that of the quantum mechanical expectation value $\langle \hat{\rho}=\frac{1}{2}(aa^\dagger+a^\dagger a)\rangle_t$. On the other hand, bosons and fermions have vastly different, and time varying, periodicities. Additionally, the quantum mechanical time evolution of the relative coherent state along periodic orbits would result in the accumulation of statistical phase factors, which for bosons and fermions are $\pm 1$ respectively. Such statistical phases have no equivalent in the classical picture presented here. Instead, the bosonic or fermionic nature of these states find expression through the symplectic form and its influence both on statistical mechanics\cite{hill}, and on classical dynamics as described in earlier section.

Further, we have already seen that what appears to be a periodic trajectory in the relative coordinate may not be periodic or even recurring in the individual particle coordinate for two-particle states. Thus the accumulated phase factors for the individual particles would not be purely geometric but depend on the applied Hamiltonian. The classical trajectories make such differences apparent whereas they would be invisible in the quantum mechanical picture. 


Lastly, we highlight a study where similar issues have been addressed in the context of effective potentials projected to the lowest Landau level. In Ref \cite{JK1}, Jain and Kivelson discuss the importance of including terms in the effective potential  of the form
\begin{align}
    \bra{R}V(x,y)\ket{R}
\end{align}
where $\ket{R}$ are coherent states in the guiding coordinates. These terms become especially important in understanding the ground state energy, dynamics as well as overall accumulated phase factors of quartic and higher order potentials in the lowest Landau level. Based on the results of this work, we argue that such results would be further modified if the states of interest included a notion of exchange statistics.  

\section{Summary and outlook}

We have built on the work in Ref. \cite{hill} concerning statistical mechanics and thermodynamics to investigate the modified classical dynamics of particles that keep a memory of their quantum statistics. This formulation is geometrical, utilizing the mathematics of K\"{a}hler manifolds, and effects of exchange statistics are encoded in the symplectic form of the classical phase space manifold. Our studies offer a generalized formulation for capturing key quantum statistical effects in a classical description of many-body dynamics in the presence of arbitrary potential landscapes.

Applying this method to scattering in one dimension we demonstrated the importance of classical statistics by comparing how fermions and bosons scatter against an inverted harmonic potential at different energies. In addition to studying their trajectories, we also showed that fermions tend to spend a smaller fraction of close to each other time as compared to bosons, and we interpreted this as a classical manifestation of the ”exclusion statistics” associated with fermions.

We then considered a two-dimensional system of a pair of vortices in the lowest Landau
level, subjected to quadratic potentials and of direct relevance to quantum Hall systems. In the presence of elliptical and saddle potentials, we observe that bosons and fermions experience an ``interaction" arising from their statistics that is absent for distinguishable particles. The most interesting effects were seen in elliptical potentials, where the trajectories of bosons and fermions showed radically different behavior, with a clear tendency for the bosons to attract each other in early stages of the simulation. During very long simulations, the trajectories tend to densely cover the regions of phase space allowed by energy conservation. This covered area differs in both size and shape between fermions and bosons. At large initial separations, when the symplectic factor for all particle species approaches unity, the difference between the trajectories vanishes. 

These studies, from the fundamental perspective one the one hand, reflect a range of possibilities for statistically-induced effects on dynamics of pairs of particles even in the simplest of potential landscapes. Extensions of our treatment to anyons are conceptually straightforward and are for future work. On the other hand, our studies are relevant to some of the most germane considerations in the recent interest that has surged in the context of anyons in the quantum Hall system. The potential landscapes studied here lie at the heart of the two primary experimental settings: the saddle potential captures the physics of the beam-splitter (collider) entailing two incoming and two out-going particles while circular and elliptical trapping potentials are key to interferometers wherein one particle encircles another\cite{b3,b4}. 

A major advantage of this classical approach is that it can easily be extended to higher
numbers of particles, thus obtaining effective classical trajectories for such many-body quantum states. We have included a brief, and preliminary, analysis of three particles in an elliptical potential and identified a parameter regime where we have numerical indications of a non-zero Lyaponov exponent, in-spite of the potential being harmonic. {An extension to many-body states presents the interesting possibility of developing a classical field theory  from the resultant many-body projected manifold, which would again retain a memory of the quantum statistics. We have not yet considered this possibility, but thank the referee for pointing it out. }
In the last section, we discussed the relationship between our results and a general geometric reformulation of quantum mechanics. in terms of K\"{a}hler manifolds, and stressed the new elements that come when statistics is included. In this context we compared our results with those from earlier simulations of the corresponding full quantum system. In conclusion, we have laid out a broadly applicable classical mechanics framework for studying many-particle dynamics that retains memory of quantum statistics, and by way of examples that would occur in a variety of situations, offered a glimpse into a rich range of statistics-dependant dynamical behavior.

\acknowledgments{We thank Predrag Cvitanovic for very valuable comments and discussions. We acknowledge the support of the National Science Foundation through Grant No. DMR- 2004825 (VS and SV).}
\bibliography{references}
\pagebreak
\appendix

\section{The classical and quantum approaches for distinguishable particles}

{In the main text of this work, we have highlighted the K\"{a}hler manifold that is obtained when a set of quantum states are projected appropriately. Therefore, the classical behavior studied in our work is obtained from the symplectic structure and resultant Poisson bracket of the projected manifold, rather than the smallness of $\hbar$. The quantum nature of the states being projected (like the non-zero commutators of their generating operators) leave indelible signatures in the dynamics that make the quantum and classical trajectories differ from each other. In this appendix, we explicitly demonstrate these signatures by considering the case of single particle coherent states for simplicity. } 

 The conventional way of studying the dynamics of a set of coherent states under a given Hamiltonian $H$ is to simply evolve them under the action of the unitary $U(t)=e^{-iHt}$. The evolution of Hermitian operators, like position or momentum,  is determined similarly as expectations over the time-evolved states. We shall call this method of obtaining trajectories the quantum approach, in contrast to the classical approach detailed in the main text. In comparing them, we specifically focus on the evolution of quadratic operators. This is because in the quantum approach, expectation values of linear operators like $\braket{\hat{x}}\sim \braket{a+a^\dagger}$ are identically zero in fermionic or bosonic states. This follows rather straightforwardly from Equation \eqref{bosferm}. Quadratic operators, on the other hand, have non zero expectation values.

Let us initially consider the simplest case of single particle coherent states $\ket{z}$. The quadratic function/potential $\rho=\frac{1}{2}(z\bar{z}+\bar{z}z)$ would classically correspond to simply $\rho=\bar{z}z$. However, quantum mechanically, it is matched to $\frac{1}{2}(aa^\dagger+a^\dagger a)$, the expectation value of which is $\rho+\frac{1}{2}$. This additional factor of $\frac{1}{2}$ has no impact when the states are first projected onto the effective classical manifold and quadratic operators are then time evolved. But it is crucial to quantum mechanical time evolution, and affects the resultant expectation values. Therefore, one can think of the two approaches as differing in the order of the operations - {\it projection onto the effective classical manifold} \ie taking expectation values and {\it time evolution}. {Such quantum mechanical effects accumulate to yield the more significant differences \sout{seen} in the trajectories of two-particle coherent states as shown in Figure \ref{mismatch1}.}

Here, we compare the trajectories derived from the classical approach, to the ones that can be extracted from the quantum mechanical time evolution of {single particle }coherent states. Specifically, we shall compare two methods of extracting classical motion
from the quantum mechanics of a set of coherent states $\{\ket {z_i}\}$:
(i) Determine an effective Lagrangian for the classical parameters $z_i$ by projecting onto an effective classical phase space, and then derive and solve the corresponding equations of motion. That is, evaluate the Lagrangian as an expectation value and then time-evolve. We will call this the classical approach. (ii) Time evolve the coherent state by the standard unitary quantum mechanical time evolution, and then determine trajectories as the expectation values of the operators corresponding to spatial coordinates. We will call this the quantum approach.

\begin{figure}
   \centering
\includegraphics[width=0.7\textwidth]{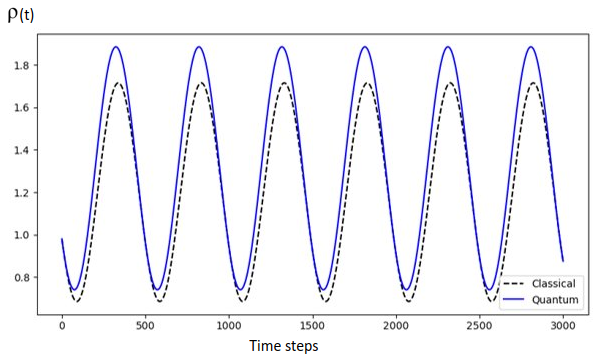}
    \caption{The equivalent of $\rho(t)=\bar{z}z$ is plotted in the classical and quantum cases. They do not match.}
   \label{mismatch}
\end{figure}

  
 It is far from obvious that these two methods will give the same trajectories.  As can be seen in Figure  \ref{mismatch}, the naive notion that the operations of ``evolving in time" and ``calculating expectation values" should commute is not true in the case of particles in the LLL moving in an elliptic potential, and this result is easily generalized to other cases.
  


  We now proceed to discover the source of the difference between the two approaches using single particle coherent states $\ket{z}$ in the lowest Landau level under the action of an elliptical potential. 

\subsection{The classical approach - project first, and then find equations of motion}
\begin{figure}
   \centering
\includegraphics[width=0.7\textwidth]{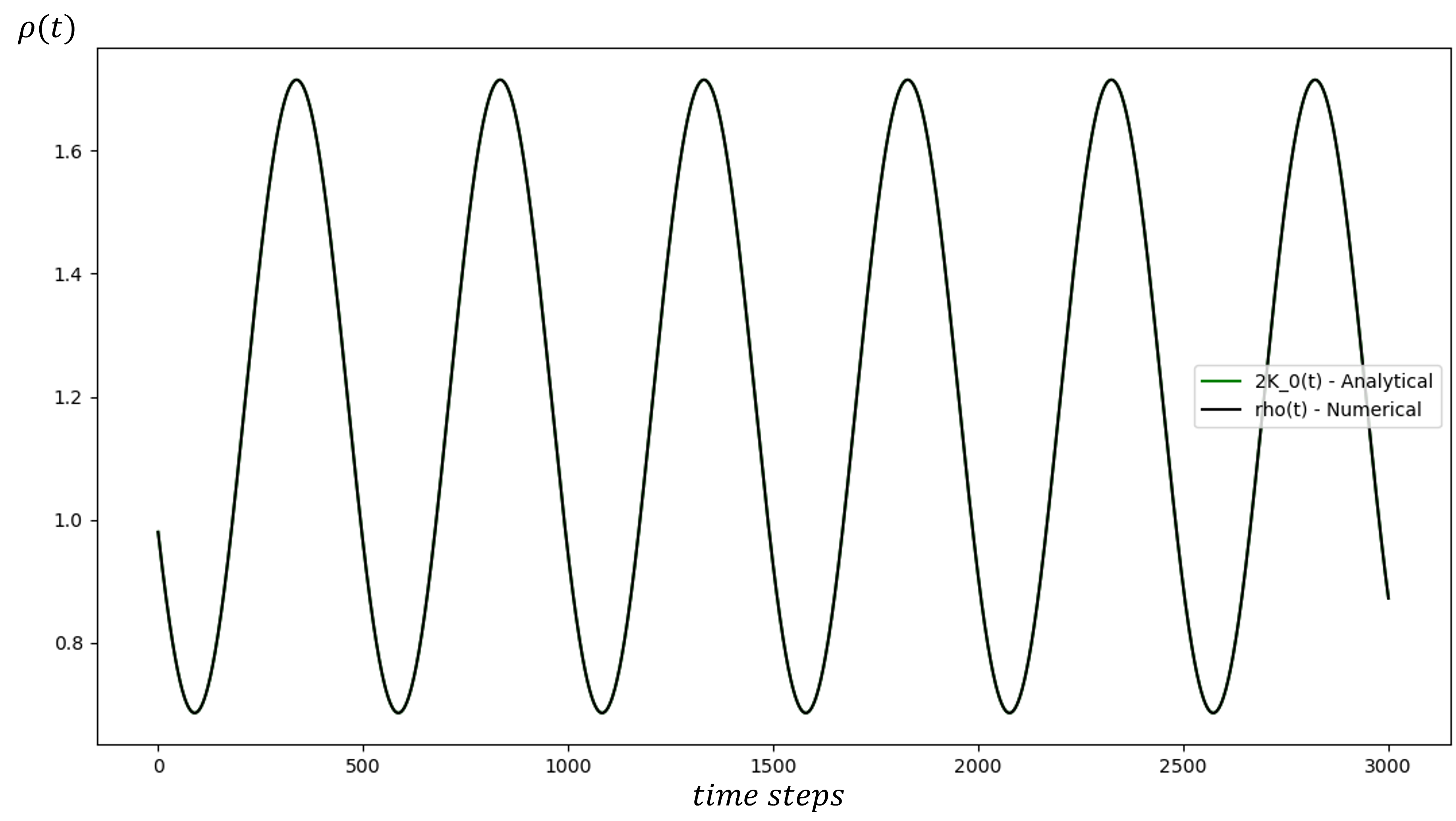}
\caption{Classically, $\rho(t)=\bar{z}z$(numerical) and $2K_0(t)$(analytical) match exactly}
\label{EOMfig}
\end{figure}
The projected classical Lagrangian for this system can be evaluated as $L=\bra{z}i\partial_t-H\ket{z}$ for an elliptical potential which yields - 
\begin{align*}
    L=\frac{i}{2}(\dot{z}\bar{z}-\dot{\bar{z}}z)-\frac{U}{4}(\bar{z}z+z\bar{z})-\frac{V}{4}(z^2+\bar{z}^2).
\end{align*}
The classical equations of motion for this Lagrangian are - 
\begin{align*}
    \dot{z}=-i\frac{U}{2}z-i\frac{V}{2}\bar{z}
\end{align*}
Equivalently, it is useful to write the equations of motion in the radial coordinates $\rho=|z|^2$ and $\theta=2\arg{z}$ as
\begin{align*}
    \dot{\rho}&=-V\rho\sin\theta\\
    \dot{\theta}&=-U-V\cos\theta.
\end{align*}
We can numerically integrate these equations to obtain $\rho(t)$, but also solve them analytically. For ease of comparison with quantum mechanical quantities in the later section, let us define the quantities: $K_c=\rho\cos\theta$, $K_s=\rho\sin\theta$ and $K_0=\rho/2$. In terms of these quantities, we can rewrite the equations of motion as - 
\begin{align}
\begin{split}
    \dot{K_c}&=UK_s\\
    \dot{K_s}&=-UK_c-2VK_0\\
    \dot{K_0}&=-\frac{V}{2}K_s.
    \end{split}
\end{align}
Let us pick an initial condition of $\rho(0)=l^2$ and $\theta(0)=\pi/2$. This gives the following solutions: 
\begin{align}
\begin{split}
    K_c(t)&=\frac{Ul^2}{\omega}\sin\omega t +\frac{UVl^2}{\omega^2}\cos\omega t -\frac{UVl^2}{\omega^2}\\
    K_s(t)&=l^2\cos\omega t -\frac{Vl^2}{\omega}\sin\omega t\\
    K_0(t)&=-\frac{Vl^2}{2\omega}\sin\omega t -\frac{V^2l^2}{2\omega^2}\cos\omega t +\frac{l^2}{2}(1+\frac{V^2}{\omega^2}).
\end{split}
\end{align}
where $\omega=\sqrt{U^2-V^2}$. The analytic and numerical solutions for $\rho(t)$ and $2K_0(t)$ are plotted in Figure  \ref{EOMfig}.
\subsection{The quantum approach - time evolve first, project later}
Writing the Hamiltonian using the angular momentum operators in the LLL  \sout{This} yields, 
\begin{align*}
    \hat{H}&=\frac{U}{4}(a^\dagger a+aa^\dagger)+\frac{V}{4}(a^2+{a^\dagger}^2)\\
    &=U\hat{K_0}+\frac{V}{2}(\hat{K}_++\hat{K}_-) \, ,
\end{align*}
where it is most natural to represent the Hamiltonian in terms of generators of quadratic potentials given by
\begin{align*}
    \hat{K}_+=\frac{{a^\dagger}^2}{2},\quad \hat{K}_-=\frac{{a}^2}{2},\quad \hat{K}_0=\frac{1}{4}(a^\dagger a + aa^\dagger)
\end{align*}
These operators generate the $\mathfrak{su}(1,1)$ Lie algebra. 

We  evaluate quantities like ${}_t\bra{z} a^\dagger a\ket{z}_t$ where $\ket{z}_t=e^{-i\hat{H}t}\ket{z}_{t=0}$, or, equivalently,  use the Ehrenfest theorem to obtain equations for how these expectation values vary in time. This is given by 
\begin{align*}
    \frac{d\langle \hat{O}\rangle}{dt}=-i\langle [\hat{O},\hat{H}]\rangle
\end{align*}
for some operator $\hat{O}$. In analogy with the classical equations, we can write 
\begin{align}
\begin{split}
    \frac{d\langle \hat{K_c}\rangle}{dt}&= \frac{d}{dt}(\langle \hat{K}_-\rangle+\langle \hat{K}_+\rangle)=U\langle \hat{K_s}\rangle\\
     \frac{d\langle \hat{K_s}\rangle}{dt}&=i\frac{d}{dt}(\langle \hat{K}_-\rangle-\langle \hat{K}_+\rangle)\\&=-U\langle \hat{K_c}\rangle-2V\langle \hat{K_0}\rangle\\
     \frac{d\langle \hat{K_0}\rangle}{dt}&=-\frac{V}{2}\langle \hat{K_s}\rangle.
     \end{split}
\end{align}
Note that these equations are exactly identical to their classical counterparts. However, the boundary conditions for the quantum case are slightly different. In particular, the quantity $\langle \hat{K_0}\rangle_{t=0}$ corresponds to $\frac{1}{2}(\rho(0)+\frac{1}{2})$ here instead of $\rho(0)/2$ in the classical case. This leads to the somewhat modified analytical solution
\begin{align}
\begin{split}
    \langle \hat{K_c}\rangle_t&=\frac{Ul^2}{\omega}\sin\omega t +\frac{UV}{\omega^2}(l^2+\frac{1}{2})\cos\omega t -\frac{UV}{\omega^2}(l^2+\frac{1}{2})\\
    \langle \hat{K_s}\rangle_t&=l^2\cos\omega t -\frac{V}{\omega}(l^2+\frac{1}{2})\sin\omega t\\
    \langle \hat{K_0}\rangle_t&=-\frac{Vl^2}{2\omega}\sin\omega t -\frac{V^2}{2\omega^2}(l^2+\frac{1}{2})\cos\omega t \\&+\frac{1}{2}(l^2+\frac{1}{2})(1+\frac{V^2}{\omega^2}).
\end{split}
\end{align}
\begin{figure}
   \centering
\includegraphics[width=0.7\textwidth]{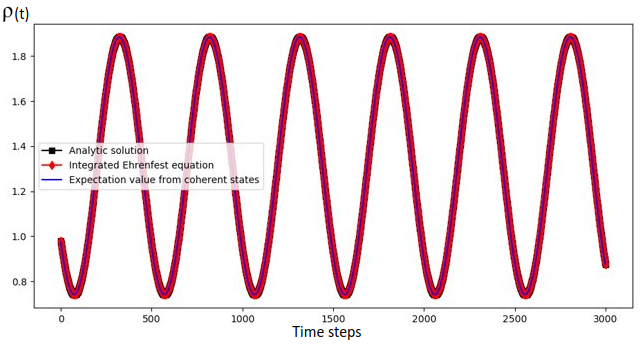}
\caption{From the quantum mechanical approach, we see that expectation values of the time evolved coherent states, the solutions of the Ehrenfest equation and the analytical solutions for the same all match. The equivalent of $\rho(t)$ is plotted in each case.}
\label{qtm}
\end{figure}
Figure  \ref{qtm} displays three different quantities that coincide - the numerical integration of the above equations to obtain $\rho(t)$, the analytically solved solution to $\langle a^\dagger a\rangle_t$ under these modified boundary conditions and the expectation value of $a^\dagger a$ calculated after explicitly time evolving the coherent state. We expect the last 2 to be identical anyway, but this serves as a way to ensure that the calculations are accurate/consistent. 
\begin{figure}
   \centering
\includegraphics[width=0.7\textwidth]{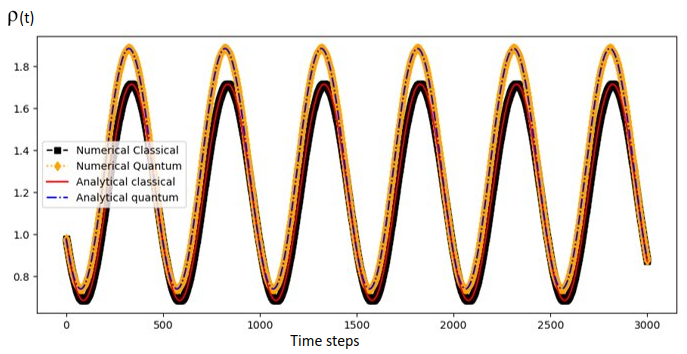}
\caption{Reproduction of Figure \ref{mismatch} with analytical solutions}
\label{repr}
\end{figure}
Next, it is straightforward to recreate Figure  \ref{mismatch} with the analytical solutions to show that we have indeed reproduced the source of the mismatch. This is done in Figure  \ref{repr}.

\end{document}